\documentclass[aps,prd,twocolumn,groupedaddress,showpacs]{revtex4}

\usepackage{graphicx}
\usepackage{graphicx}
\usepackage{amssymb}
\usepackage{amsmath}
\usepackage{color}
\newcommand{\be}{\begin{equation}}
\newcommand{\ee}{\end{equation}}
 
 \definecolor{BrickRed}{cmyk}{0,0.89,0.94,0.28}%%%PANTONE 1805
\definecolor{MidnightBlue}{cmyk}{0.98,0.13,0,0.43}%%%PANTONE 302
\definecolor{DarkGreen}{rgb}{0,0.7,0.1}

\begin{document}

\title{Beyond-proximity-force-approximation Casimir force between two  spheres at finite temperature II: \\ plasma  versus Drude modeling, grounded versus isolated spheres}

%\date{\today}

\author{Giuseppe Bimonte}

\affiliation{ Dipartimento di Fisica E. Pancini, Universit\`{a} di
Napoli Federico II, Complesso Universitario
di Monte S. Angelo,  Via Cintia, I-80126 Napoli, Italy}
\affiliation{INFN Sezione di Napoli, I-80126 Napoli, Italy}

% e-mail addresses: one for each author, in the same order as the authors
\email{giuseppe.bimonte@na.infn.it}

\begin{abstract}

A recent experiment  [J. L. Garrett, D. A. T. Somers, and J. N. Munday, Phys. Rev. Lett {\bf 120}, 040401 (2018)] measured for the first time the gradient of the Casimir force  between two gold spheres in vacuum at room temperature, and placed a bound on the magnitude of the deviation of the measured force from  the proximity force approximation (PFA). The present work extends a previous theoretical analysis of this experiment [G. Bimonte, Phys. Rev. D {\bf 97}, 085011 (2018)], by analyzing  in detail
how the magnitude of the deviation from PFA  is affected by the  inclusion  or neglect of ohmic dissipation at zero frequency, a much debated issue in the current Casimir literature,  which goes by  the name of the Drude vs plasma controversy. We analyze as well the effect of connecting the conductors to charge reservoirs, which is the standard configuration used in Casimir experiments.   We describe a simple and effective decimation procedure, allowing for a faster computation of the Casimir force for large aspect ratios of the system.

\end{abstract}


\maketitle

\section{Introduction}
\label{sec:intro}

In the last two decades  many experiments by different groups across the world have been carried out to  measure the Casimir force \cite{Casimir48}, and  huge efforts have been made by skilled experimentalists to achieve an ever increasing precision (for  reviews   see Refs.\cite{book1,parse,book2,buhmann,woods,mehran}). There are distinct good reasons behind the quest for precision in Casimir experiments. One one hand, precise measurements challenge our fundamental understanding  of subtle properties of dispersion forces such as their non-additivity and shape dependence,  as well as their dependence on material properties of the bodies, like temperature and the optical characteristics of the surfaces \cite{parse,book2}. On the other hand, controlling the  Casimir force is essential in current searches of  non-newtonian gravity in the sub-micron range \cite{decca6,decca3}. Since the  sensitivity of current devices for measuring small forces is not yet sufficient to observe the gravitational force at these small distances, the experiments  just place bounds on the the magnitude of possible deviations from Newton's law, whose strength depends crucially on the theoretical uncertainty on the magnitude of the much stronger Casimir force between the test bodies.       

In parallel with the experimental investigations and motivated by them, intense efforts have been  made by theoreticians to improve the precision of computations of the Casimir force. Until not long ago nobody knew how to exactly compute the Casimir force in non-planar setups, like for example the standard experimental configuration of a sphere and plate.  The available theoretical arsenal was rather meager and it basically consisted of the old-fashioned proximity force approximation (PFA) introduced in the 30's of the 19th century by Derjaguin \cite{Derjaguin} to deal with short-range forces between two curved surfaces. Within this approximation, the Casimir force between two curved surfaces is expressed as the average over the local separation of the Casimir pressure between two plane-parallel dielectric slabs, whose expression was derived by Lifshitz over 60 years ago \cite{lifs}.  Corrections to PFA were generically expected to be the order of $a/R$, with $a$ the minimum separation and $R$ the characteristic radius of curvature of the surfaces, but nobody could be sure of that. In order to obtain  forces that are large enough to be measured precisely, experiments always use systems with large aspect ratios $R/a$,  and so it has always been given for granted that deviations from  PFA are negligible.  Agreement with experimental data has been indeed reported in all cases, apart from one very surprising fact:  the most precise  room temperature sphere-plate experiments   utilizing gold-coated surfaces, carried out by different groups, showed that agreement with theoretical predictions is only possible if the optical data of gold are extrapolated towards zero frequency by  the dissipantionless  plasma model of infrared optics, while the better motivated  dissipative Drude model is ruled out. This has come to be known in the Casimir community as the Drude versus plasma conundrum. For a review of this problem, see \cite{book2}  (see also the recent experiment \cite{deccamag} and references therein).

Despite its plausibility, the assumption underlying the theoretical analysis of the experiments, that deviations from PFA would be too small to fill the gap between the plasma and the Drude models, could not be taken for granted and remained open to challenge.
The situation changed only  around 2005, when an extension of early results by Balian and Duplantier \cite{Balian} and Langbein \cite{langbein} led to the discovery of an {\it exact} scattering formula \cite{sca1,sca2,kenneth} for the Casimir free energy of two compact dielectric bodies    
of any shapes. The scattering formula provides a representation of the Casimir free energy in the form of a functional determinant involving the T-matrices of the two bodies, and the translation matrices that serve to express the multipole basis relative to  either body in terms of the multipole basis of the other.  The scattering formula led to rapid progress. With its help it  was proved rigorously  that the PFA is indeed exact in the asymptotic limit $a/R \rightarrow 0$ for both the sphere-plate and sphere-cylinder geometries\cite{bordag}, it was possible to work out a systematic perturbative expansion to compute the effect of small corrugations \cite{paulo},  and to analytically compute the {\it leading} correction beyond PFA in a number of distinct geometries \cite{bordag,fosco1,bordagteo,bimonte1,fosco2,fosco3,bimonte2},  confirming that they are indeed of order $a/R$, as it had been assumed before.  

The reader may wonder at this point why one should be content with knowing just the leading order correction to PFA, rather than trying to directly compute the scattering formula numerically, in order to obtain virtually exact values for the Casimir force.  Unfortunately, even when the T-matrix is known exactly (as it is the case for dielectric spheres and plates) computing the scattering formula numerically is not at all an easy task, even with the help of the powerful  computers that are available today. The problem is that  the multipole order for which convergence is achieved scales like  $R/a$, and therefore  for typical  experimental aspect ratios $R/a$ of the order of $10^{3}$ or larger, tens of thousands of multipoles are needed. For over ten years after the discovery of the scattering formula, such a large number of multipoles has been out of reach, and numerical simulations were restricted to small aspect ratios $R/a$ less than one hundred \cite{antoine1,antoine2}.  Only last year \cite{gert} significant numerical improvements using state-of-the-art  algorithms for the computation of determinants of hierarchical off-diagonal low-rank matrices, allowed to handle for the first time tens of thousands of multipoles,  permitting to compute deviations from PFA  in the sphere-plate configuration for experimentally relevant aspect ratios around one thousand.  The computed deviations from PFA were compared to the experimental bound  obtained in the experiment \cite{ricardo},  and it was concluded that the Drude model  is indeed in better agreement with the data than the plasma model.

The numerical algorithms used in \cite{gert} are rather sophisticated and  are not easy to implement for non experts. At about the same time when  \cite{gert}  appeared, an alternative semianalytic approach  was presented in \cite{bimonteprecise}, which allows to reach the same  level of precision  with a far smaller numerical effort. The latter approach combines the leading-order correction to PFA for positive Matsubara modes,  which can be computed by means of the derivative expansion (DE)  \cite{fosco1,bimonte1,fosco2,fosco3,bimonte2}, with the exact sphere-plate formula for the zero-frequency contribution  that was worked out in \cite{bimonteex1} for two metallic bodies modeled as Drude conductors. The effectiveness of this approach in providing a remarkably precise expression of the Casimir force for all separations,  hinges on the fact that for positive Matsubara frequencies the electromagnetic field effectively behaves as a massive field, with a mass proportional to the temperature $T$ and to the Matsubara discrete index $n$.   As a result of this feature,  the Casimir interaction is short-ranged for positive Matsubara frequencies,  and this in turn renders the DE very precise. A drawback of the approach presented in \cite{bimonteprecise}  is that in the simple version discussed there, it cannot be applied to the plasma model, because no exact formula exists for its zero-frequency contribution for transverse electric (TE) polarization.    

The vast majority of experimental and theoretical  investigations of the Casimir effect focused in the past on the sphere-plate geometry. Very recently, however, a new experiment \cite{garrett} has measured for the first time the (gradient of the) Casimir force between two gold-coated spheres. The data were found to be in good agreement with   theoretical predictions based on the standard PFA.  Following a procedure analogous to that of the sphere-plate experiment  by the IUPUI group \cite{ricardo}, the authors of \cite{garrett}  used  nine sphere-sphere and three sphere-plate setups of different radii to obtain an experimental bound on the magnitude of deviations from the PFA. 

Prior to this experiment, the electromagnetic two-sphere problem received little attention in the literature: in \cite{pablo} the Casimir energy of two metallic spheres was studied in the limit of large separations, while in \cite{teo2} it was demonstrated that  the small-distance limit of the two-sphere scattering formula indeed reproduces the PFA. In particular, there were no published works  devoted to computing  beyond PFA corrections for this system, that could be directly compared with the experimental data of \cite{garrett}.   Motivated by the lack of theoretical predictions, the computation of the force-gradient (the quantity actually measured in the experiment)  was undertaken in \cite{bimonte2sphere}, following the same approach that was used  in \cite{bimonteprecise} to study the  sphere-plate system.  In \cite{bimonte2sphere}  analysis was restricted to the  experimental configuration of two {\it grounded} gold spheres, modeled as Drude conductors. This case is well suited to the scheme of \cite{bimonteprecise}, because  for two grounded Drude conductors the electromagnetic zero-frequency Matsubara contribution  coincides with the corresponding contribution for two Dirichlet-spheres, for which an exact formula was worked out in \cite{bimonteex1} exploiting conformal invariance (the same formula has been later derived directly from the scattering formula using a similarity transformation in \cite{teo2s}). The deviation from PFA obtained in \cite{bimonte2sphere} were found to be consistent with the (rather loose) experimental bound placed in \cite{garrett}.

The present work extends \cite{bimonte2sphere} in two respects: in the first place, corrections to PFA are presented here for both  the Casimir force and its gradient. Second, and more importantly, deviations from PFA are computed here for  both the plasma model, as well as for the Drude model. We also address in detail the effect of grounding the spheres, by separately considering  both cases of {\it grounded} or  {\it isolated} conductors. We underline that in Casimir experiments with conducting surfaces, the conductors are always connected to charge reservoirs in order to get rid of possible charges that may be present on the surfaces, and/or to apply a bias potential to compensate for differences among the work functions of the surfaces \cite{deccapat}. Despite this, practically all theoretical studies implicitly assume that the conductors as isolated, by using the expression of the $T$-matrices (the Mie coefficients in the case of spheres) that actually describe isolated conductors. The question of grounding or not the conductors is explicitly discussed  in very few works  \cite{bimonteex1,foscogr}.  In the current experimental situation, the distinction between  grounded or isolated conductors is more a matter of principle than a necessity, since 
the difference between the two models  manifests itself only at the level of beyond PFA deviations, which are too small to be detected with  current apparatus.
    
The analysis of  the new models studied in this paper, i.e. the plasma  model and the Drude model with isolated conductors (which coincides with what is  usually understood as Drude model, with no other qualifications, in the Casimir literature) is considerably more difficult than the model of grounded Drude conductors studied in \cite{bimonte2sphere}. The positive Matsubara modes present no particular difficulty, and  can be accounted for by means of the DE.  However, the zero-frequency contribution is problematic, since no exact solution exists for either the  Drude or the plasma models. One could imagine using the DE to estimate this contribution as well, of course. Unfortunately, within the plasma model the DE is known to fail  for the zero-frequency  TE mode \cite{fosco4, bimonteHT}, and even for the Drude model, where it is applicable,  it is expected to be less precise for moderate values of $R/a$, due to the massless character of the transverse electric (TM) mode for zero frequency. So, a different route is necessary.  The Drude model is easier, because  its   zero-frequency contribution can  be computed numerically, quickly and with high precision, by expressing the exact scattering formula in a  {\it bispherical} multipole basis \cite{bimonteex1}. The huge advantage offered by the latter basis, as contrasted with the standard spherical basis (in which the origins of the multipoles are placed  at the centers of the two spheres)  is that it ensures  a much faster convergence rate, since the required (bispherical) multipole order scales only like $\sqrt{R/a}$, rather than $R/a$. This implies that for aspect ratios $R/a$ of the order of, say, $10^3$ just 100 bispherical multipoles are sufficient, instead of ten thousands!  Bispherical coordinates are unfortunately of no help with the plasma model, because the  corresponding boundary conditions
for zero-frequency TE polarization  cannot be expressed in a simple manner in this coordinate system (while the TM contribution is identical to the Drude model). To handle the TE zero-frequency contribution it is necessary to resort to the conventional spherical multipole basis. To cope with the slow convergence of the latter for large aspect ratios, a very simple {\it decimation} procedure of the scattering formula can be devised, that allows to reduce by a significant factor the size of the involved matrices without significantly  jeopardising  precision. It turns out  that for gold  the plasma model deviations from PFA can be accurately reproduced  by just taking to infinity the plasma frequency, when evaluating the zero-frequency contribution. In this limit, the zero-frequency  plasma model coincides with the perfect-conductor model,  and in a previous work \cite{bimonteHT} it was shown that the latter model can  be computed efficiently in bispherical coordinates.

The plan of the paper is as follows: in Sec. II we describe the two-sphere system, review the general scattering formula, and we show how  the DE can be used to compute the contribution of the positive Matsubara modes. In Sec. III we discuss the zero-frequency contributions, within the Drude and the plasma models, and we analyze the effect of grounding or not the conductors. In Sec. IV we describe a simple decimation procedure for the scattering formula, which allows for a faster computation of the Casimir force at zero frequency for large aspect ratios of the system. In Sec. V we present the results of our numerical computations, while in Sec. VI we present our conclusions.

\section{Casimir interaction of two spheres}

We consider a system of two spheres of respective radii $R_1$ and $R_2$ placed in vacuum, and we let $a$ the tip-to-tip distance (see Fig. 1). We adopt here the same parametrization of the two sphere system that was used in \cite{bimonte2sphere}. According to this parametrization the geometry of the system is characterized by the effective radius  $\tilde R=R_1 R_2/(R_1+R_2)$ and by the dimensionless parameter $u={\tilde R}^2/(R_1 R_2)$. 

\begin{figure}
\includegraphics[width=.9\columnwidth]{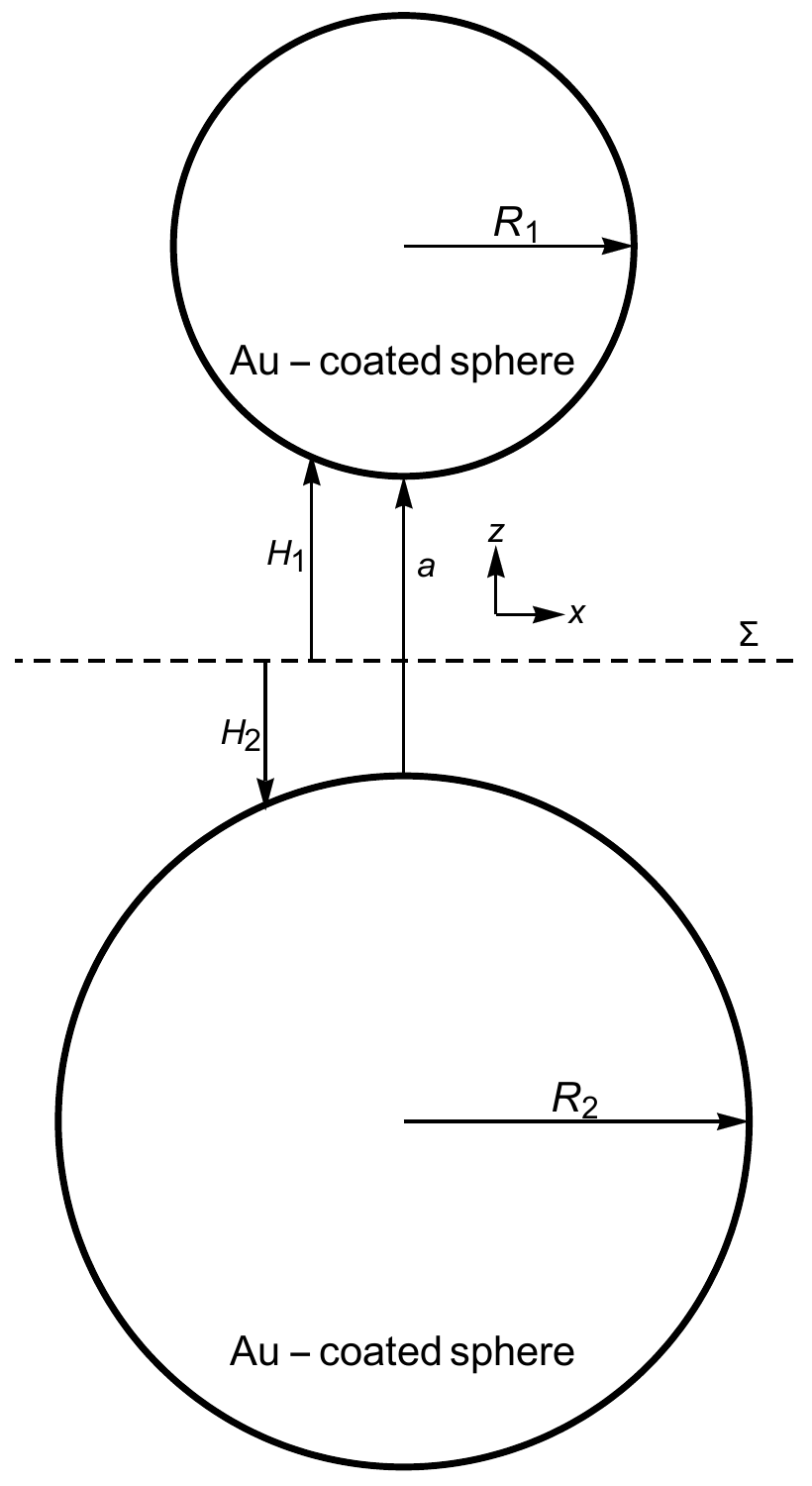}% Here is how to import EPS art
\caption{\label{setup}  The sphere-sphere Casimir setup. The sphere-sphere geometry is characterized by the {\it effective} radius ${\tilde R}=R_1 R_2/(R_1+R_2)$ and the dimensionless parameter $u={\tilde R}^2/R_1 R_2$}
\end{figure} 
The exact Casimir free energy ${\cal F}$ of a two-body  system is provided by the scattering formula. We recall that the  general form of this formula \cite{sca1,sca2,kenneth} for  two objects of any shape (denoted as 1 and 2) in vacuum is:
\be
{\cal F}=k_B T \sum_{n \ge 0}\,\!\!' \,{\rm Tr \ln} [1-{\hat N}({\rm i} \,\xi_n) ]\;,\label{freeen0}
\ee
where the operator ${\hat N}({\rm i} \,\xi_n)$ is
\be
{\hat N}={\hat T}^{(2)}{\hat U}{\hat T}^{(1)}{\hat U}\;.\label{Mmat}
\ee
Here $k_B$ is Boltzmann's constant, $T$ is the temperature,   $\xi_n= 2 \pi n k_B T/\hbar$ are the (imaginary) Matsubara frequencies, and the prime in the sum indicates that the $n=0$ term is taken with weight 1/2.  In Eq. (\ref{Mmat}), ${\hat T}^{(j)}$ denotes the T-operator of object $j$, evaluated for imaginary frequency ${\rm i} \,\xi_n$, and ${\hat U}$ is the translation operator that  serves to transform the chosen basis of outgoing fields relative to  one  of the two objects into the basis of  ingoing fields relative to  the other object. 
In the specific case of two spheres, the scattering formula is usually expressed in terms of two sets of spherical multipoles, whose origins are placed at the centers of the two spheres. In this basis, the matrix elements of the T-operators of the two (isolated) spheres coincide with the respective Mie coefficients. The  explicit expression of the matrix elements of the translation operators   ${\hat U} $  can be found  in Refs.\cite{rahi,teo2}, and shall not be written here for brevity.  Symmetry of the two sphere system under rotations around the $z$-axis passing through their centers, as well as under reflections in the $(x,y)$ plane, allows to express the scattering formula as a sum over non-negative eigenvalues $m$ of the $z$-component of the angular momentum:
\be
{\cal F}=2 k_B T \sum_{n \ge 0}\,\!\!' \sum_{m=0}^{\infty}\,\!\!' \,{\rm Tr \ln} [1-{\hat N}({\rm i} \,\xi_n;m) ]\;,\label{freeen}
\ee
where ${\hat N}({\rm i} \,\xi_n;m) $ denotes the restriction of the operator ${\hat N}$ to the subspace of multipoles  with azimuthal number $m$, and the prime in the sum over $m$ indicates that the $m=0$ term has to be taken with weight 1/2.  
 The trace ${\rm Tr}$  in Eq. (\ref{freeen}) is taken  over the spherical multipoles index $l$ as well as  over the polarizations $\alpha={\rm TE,TM}$:
\be
{\rm Tr}=  \sum_{l=|m|}^{\infty} {\rm tr}\;,
\ee  
where ${\rm tr}$ denotes the trace over $\alpha$. The Casimir force $F=-  {\cal F}'$ and its gradient $F'$  are obtained by taking  derivatives of Eq. (\ref{freeen}) with respect to the separation $a$. The explicit scattering formula for the Casimir force is:
\be
{F}= 2 k_B T \sum_{n \ge 0}\,\!\!' \sum_{m=0}^{\infty}\,\!\!' \,{\rm Tr}\left[\frac{\partial_a{\hat N}}{1-{\hat N}} \right] \;,\label{scattF}
\ee
while the formula for the force-gradient is:
\be
{F'}= 2 k_B T \sum_{n \ge 0}\,\!\!' \sum_{m=0}^{\infty}\,\!\!'  \,{\rm Tr}\left[\frac{\partial^2_a{\hat N}}{1-{\hat N}}+\left(\frac{\partial_a{\hat N}}{1-{\hat N}} \right)^2\right] \;,\label{scattF'}
\ee
where $\partial_a$ denotes a derivative with respect to  the separation $a$.
Starting from the scattering formula, it is possible to show \cite{teo2} that in the limit of large aspect ratios ${\tilde R}/a \rightarrow \infty$ the Casimir force $F$  approaches the PFA:   
\be
{F}^{(\rm PFA)}(a,R_1,R_2)= 2 \pi {\tilde R}\; {\cal F}^{(\rm pp)}(a)\;,\label{PFA}
\ee
where  ${\cal F}^{(\rm pp)}(a)$ denotes the famous  Lifshitz formula \cite{lifs} providing the unit-area Casimir free energy for two plane-parallel slabs respectively made of the same materials as the two spheres:
\begin{eqnarray}
&&
{\cal F}^{(\rm pp)}(a,T)=\frac{k_BT}{2\pi} \sum_{n \ge 0}\;\!\!'
 \int_{0}^{\infty}
k_{\bot}dk_{\bot} \nonumber
\\
&&\!\!\times
\!\!\!\sum\limits_{\alpha={\rm
TE,TM}}\ln\left[1-
 { r_{\alpha}^{(1)}({\rm i}\,\xi_n,k_{\bot})\, r_{\alpha}^{(2)}({\rm i}\,\xi_n,k_{\bot})}e^{-2aq_n}\right].\label{PBeq3}
\end{eqnarray}
In this equation, $k_{\bot}$ is the in-plane momentum, $r_{\alpha}^{(i)}({\rm i}\,\xi_n,k_{\bot})$ denotes the Fresnel reflection coefficient  of  the $ i$-slab (in our case both slabs are made of gold) for polarization $\alpha={\rm TE, TM}$, evaluated for the imaginary frequency $\omega={\rm i}\, \xi_n$ and 
$q_n=\sqrt{\xi_n^2/c^2+k_{\bot}^2}$. 

The PFA formula for the force gradient is obtained by taking a derivative of Eq. (\ref{PFA})  with respect to  the separation $a$:
\be
{F'}^{(\rm PFA)}(a,R_1,R_2)= -2 \pi {\tilde R}\; {F}^{(\rm pp)}(a)\;,\label{PFA2}
\ee
 where ${F}^{(\rm pp)}=-\partial {\cal F}^{(\rm pp)}/\partial a$ is the Casimir pressure between two plane-parallel slabs. 

Our aim is to go beyond the PFA,  by performing a precise computation of both the Casimir force and force gradient.  
Our strategy to obtain precise values of the Casimir force and its gradient is similar to the one adopted in \cite{bimonteprecise,bimonte2sphere}. We start by separating in Eq. (\ref{freeen}) the classical  term ${\cal F}_{ n=0}$ from the contribution ${\cal F}_{n>0}$ of the positive Matsubara modes, and we accordingly split the free energy as:
\be
{\cal F}={\cal F}_{n=0}+{\cal F}_{n>0}\;.
\ee
The corresponding decompositions of the force $F$ and force gradient $F'$ are:
\be
{F}={F}_{n=0}+{F}_{n>0}\;,\label{decomF}
\ee
\be
{F'}={F}'_{n=0}+{F}'_{n>0}\;.\label{decomF2}
\ee
The zero-frequency term of the scattering formula is associated with {\it classical} (as opposed to quantum) thermal fluctuations of the electromagnetic field, and for this reason is it known in the literature as the classical term. The latter provides the dominant contribution to the Casimir interaction in the high-temperature limit $a/\lambda_T \gg 1$, where  $\lambda_T=\hbar c/(2 \pi k_B T)$ ($\lambda_T=1.2$ microns at room temperature). In the next section we shall show how to compute the classical term for several distinct models of a conductor that are frequently considered in the literature. In this section, we focus our attention on the contribution of the positive Matsubara modes.

In \cite{bimonte2sphere} it was shown that the contribution of the positive Matsubara modes  can be accurately estimated by the DE   \cite{fosco1,bimonte1,fosco2,fosco3,bimonte2}. Let us briefly recall the arguments providing support to this claim.  Compared to the classical term, positive Matsubara modes contribute significantly to the Casimir force only for separations that are not too large compared with $\lambda_T$, and are in fact dominant for $a/\lambda_T \ll 1$. This implies that for all separations for which the positive Matsubara modes matter,  the condition $a/{\tilde R}\ll 1$ is always satisfied, provided that,  as it always happens in practice, the radii of the spheres are both much larger than $\lambda_T$.  Another important thing to bear in mind is that for positive Matsubara frequencies $\xi_n$, the electromagnetic field is effectively massive, the effective  mass being proportional to $\hbar \xi_n/c^2$. The massive character of the positive modes  implies that the corresponding Casimir interaction satisfies the {\it locality} requirements, that ensure existence of the DE \cite{fosco1,bimonte1,fosco2,fosco3,bimonte2}.  In  \cite{bimonte2sphere},  using the DE,  the following  formula for ${F}'_{n>0}$ was obtained: 
\be
{ F}'_{n>0}= -2 \pi {\tilde R} {F}^{(\rm pp)}_{n>0}(a) \left[1- \left(\tilde{\theta}(a)+u \,{\tilde \kappa}(a) \right) \frac{a}{\tilde R}+o(a/{\tilde R}) \right]\;, \label{DEformula1}
\ee
where the coefficients $\tilde \theta(a)$ and ${\tilde \kappa}(a)$ \footnote{The coefficient ${\tilde \kappa}(a)$ was denoted as $\kappa$ in \cite{bimonte2sphere}.} are
\begin{eqnarray}
{\tilde \theta}&=&  \frac{{{\cal F}}^{(\rm pp)}_{n>0}(a) -2 \alpha_{n>0}(a)}{a {F}^{(\rm pp)}_{n>0}(a)}\;,\label{thetacoe1}\\
{\tilde \kappa}(a) &=&  1-2 \frac{{ {\cal F}}^{(\rm pp)}_{n>0}(a)}{ a { F}^{(\rm pp)}_{n>0}(a)} \;.\label{kappacoe1}
\end{eqnarray}
In the above Equations, $ {\cal F}^{(\rm pp)}_{n>0}$ denotes the contribution of the positive Matsubara modes to Lifshitz formula Eq. (\ref{PBeq3}), and ${F}^{(\rm pp)}_{n>0}=-\partial {\cal F}^{(\rm pp)}_{n>0}/\partial a$ is the corresponding pressure.  The first term between the square brackets on the rhs of Eq. (\ref{DEformula1}) coincides with the PFA Eq.(\ref{PFA2}) (restricted of course to positive Matsubara modes), while its second term provides the {\it leading} correction beyond PFA. An essential ingredient of Eqs. (\ref{thetacoe1}) and (\ref{kappacoe1}) is the coefficient $\alpha_{n>0}(a)$, that can be extracted from the Green function ${\tilde G}^{(2)}(k;a)$
of  the  perturbative expansion of ${\cal F}_{n>0}$, to second order in the amplitude of a small deformation of one of the surfaces around the plane-parallel  geometry.  More precisely,  $\alpha_{n>0}(a)$  is proportional to the coefficient of $k_{\perp}^2$ in the small-momentum Taylor expansion of   ${\tilde G}^{(2)}(k;a)$ (see \cite{bimonte2sphere} for details). It is the  existence of the latter Taylor expansion which is ensured by the locality properties satisfied by the Casimir interaction, for positive Matsubara frequencies.

The corresponding  formula  for the Casimir force $F_{n>0}$ can be obtained by integrating Eq. (\ref{DEformula1}) with respect to the separation $a$:
\be
{ F}_{n>0}= 2 \pi {\tilde R} {\cal F}^{(\rm pp)}_{n>0}(a) \left[1- \left({\theta}(a)+u \,{ \kappa}(a) \right) \frac{a}{\tilde R}+o(a/{\tilde R}) \right]\;, \label{DEformula2}
\ee
 where the coefficients  $\theta(a)$ and $ \kappa(a)$ are
\begin{eqnarray}
{\theta}&=&  \frac{2 \psi_{n>0}(a)-{{\cal G}}^{(\rm pp)}_{n>0}(a) }{a {\cal F}^{(\rm pp)}_{n>0}(a)}\;,\label{thetacoe2}\\
{\kappa}(a) &=&  1+\frac{{ {\cal G}}^{(\rm pp)}_{n>0}(a)}{ a {\cal  F}^{(\rm pp)}_{n>0}(a)} \;.\label{kappacoe2}
\end{eqnarray}
In these equations, the coefficient $\psi_{n>0}(a)$  coincides with the integral of $\alpha_{n>0}(a)$:
\be
\psi_{n>0} =-\int_a^{\infty} d x \,\alpha_{n>0}(x)\;,
\ee  
while ${\cal G}^{(\rm pp)}_{n>0}(a)$ coincides with the integral of $ {\cal F}^{(\rm pp)}_{n>0}(a)$ with respect to the separation $a$:
\begin{widetext}
\be
{\cal G}^{(\rm pp)}_{n>0}(a,T)=-\int_a^{\infty} dx \,  {\cal F}^{(\rm pp)}_{n>0}(x)=\frac{k_BT}{4\pi} \sum_{n > 0} \int_{0}^{\infty}
\frac{k_{\bot}dk_{\bot}}{q_n}
\sum\limits_{\alpha={\rm TE,TM}}{\rm Li}_2 \left[
 r_{\alpha}^{(1)}({\rm i}\,\xi_n,k_{\bot})\, r_{\alpha}^{(2)}({\rm i}\,\xi_n,k_{\bot})\, e^{-2aq_n}\right],
\ee
\end{widetext}
 where ${\rm Li}_s(x)=\sum_{k=1}^{\infty} x^k/k^s$ denotes the polylogarithm function. Equations (\ref{DEformula1}) and (\ref{DEformula2}) show  that within the PFA, the forces $F_{ n>0}$ and $F'_{n>0}$ depend only on the effective radius ${\tilde R}$, while the leading corrections beyond PFA do depend also on the ratio of the radii $R_1$ and $R_2$ via the parameter $u$. 

The expressions of the coefficients  $\alpha_{n>0}(a)$  and $\psi_{n>0}(a)$ are too long to be reported here. We just note that both have expressions of the form $\alpha_{n>0}/\psi_{n>0}= \sum_{n>0} \int k_{\bot}dk_{\bot} g_{\alpha/\psi} ( k_{\bot}, \epsilon({\rm i} \xi_n);a) $, where $g_{\alpha/\psi} (k_{\bot}, \epsilon(i \xi_n);a)$ are certain functions of the in-plane momentum $k_{\bot}$, of the permittiviestes $\epsilon(i \xi_n)$ and of the separation $a$, that can be easily computed numerically. The values of the coefficients $\theta(a)$, ${\kappa (a)}$, ${\tilde \theta}(a)$ and $\tilde \kappa(a)$
were computed using the tabulated optical data for gold \cite{palik}, suitably extrapolated towards zero frequency using either  the Drude prescription (with Drude parameters $\omega_p=9$ eV/$\hbar$ and $\gamma=0.035$ eV/$\hbar$) or  the plasma prescription.  The weighted dispersion relation described in \cite{bimonteKK} was used to suppress the influence of the extrapolation on the obtained values of the permittivity for positive Matsubara frequencies.  
\begin{widetext}
\label{tab.1}
\begin{center}
\begin{table*}
%\begin{tabular}{l l l l l l l l l l l | l  l} \hline
\begin{tabular}{ccccccccccccc} \hline
$a (\mu m)$\;\; &0.10& 0.15 & 0.2  \;\;& 0.25\;\;\;&\; 0.3\;\; &0.35 \;\;& 0.4 \;\; & 0.45 & 0.5\;\;& 0.55 & 0.6\;& 0.65  \\ \hline \hline
${\theta}$\;\;& 0.717  &0.694 & 0.664\;\; &0.636 &\; 0.609 \;\;&0.584 & 0.561 \;&0.540 & 0.520 &0.502 & 0.484& 0.468    \\  \hline
${\kappa}$\;\;&0.440 &0.471 & 0.496\;\; &0.515 &\; 0.532 \;\;&0.546 & 0.559\;\;&0.571  &\; 0.583\;\;&0.593 &\; 0.603\;\; &\;\;0.613 \;\;    \\ \hline \\
 \hline
$a (\mu m)$\; &0.70&0.75 & 0.8 & 0.85 & 0.9& 0.95&  1 &1.2&1.4 &1.6 & 1.8 & 2  \\ \hline \hline
${\theta}$ \;\; &0.453 &0.439  &0.425 &0.412 &0.400&0.389&0.378 &0.340&0.307&0.279& 0.256 &0.237 \\  \hline
${ \kappa}\;\;$&0.622&\;0.630 \;&0.639 &0.647&0.655&0.662& 0.669\;\;&0.696&0.719&0.739&0.757&0.774 \\ \hline  \hline
\end{tabular}
\caption{Values of the coefficients ${\theta}$ and ${ \kappa}$ for Au at room temperature (Drude prescription).}
\end{table*}
\end{center}
\label{tab.2}
\begin{center}
\begin{table*}
%\begin{tabular}{l l l l l l l l l l l | l  l} \hline
\begin{tabular}{ccccccccccccc} \hline
$a (\mu m)$\;\; &0.10& 0.15 & 0.2  \;\;& 0.25\;\;\;& 0.3\;\; &0.35 \;\;& 0.4 \;\; & 0.45 & 0.5\;\;& 0.55 & 0.6\;\; & 0.65  \\ \hline \hline
${\tilde \theta}$\;\;& 0.456  &0.4715 & 0.470\;\; &0.463 & 0.454 \;\;&0.4445 & 0.435 \;&0.425 & 0.415 &0.4055 & 0.396& 0.387    \\  \hline
${\tilde \kappa}$\;\;&0.245 &0.270 &\; 0.289\;\; & 0.305 & 0.319 \;&0.331 & 0.342\;\;&0.353  &\; 0.362\;&0.371 & \;0.380\;&\;\;0.389 \;\;    \\ \hline \\
 \hline
$a (\mu m)$\; &0.70&0.75 & 0.8 & 0.85 & 0.9& 0.95&  1 &1.2&1.4 &1.6 & 1.8 & 2  \\ \hline \hline
${\tilde \theta}$ \;\; &0.379 &0.370  &0.362 &0.3545 &0.347&0.3395&0.332 &0.306&0.282&0.261& 0.242 &0.225 \\  \hline
${\tilde \kappa}\;\;$&0.397&\;0.405 \;&0.413 &0.421&0.429&0.437&\; 0.444\;\;&0.474&0.502&0.529&0.554&0.578\\ \hline  \hline
\end{tabular}
\caption{Values of the coefficients ${\tilde \theta}$ and $\tilde \kappa$ for Au at room temperature (Drude prescription).}
\end{table*}
\end{center}
\end{widetext}
The Drude values for the two pairs of coefficients $( \theta(a),{ \kappa}(a))$ and $({\tilde \theta}(a),{\tilde \kappa(a)})$ are listed,  for several values of the separation $a$, in Tables I and  II, respectively, while the corresponding values for the plasma prescription   are  listed in Tables III and  IV, respectively.  
As it can be seen by comparing Tables I and  II with Tables III  and IV, respectively, the  Drude and plasma prescriptions lead to nearly identical values of the corresponding coefficients. 
\begin{widetext}
\label{tab.3}
\begin{center}
\begin{table*}
%\begin{tabular}{l l l l l l l l l l l | l  l} \hline
\begin{tabular}{ccccccccccccc} \hline
$a (\mu m)$\;\; &0.10& 0.15 & 0.2  \;\;& 0.25\;\;\;&\; 0.3\;\; &0.35 \;\;& 0.4 \;\; & 0.45 & 0.5\;\;& 0.55 & 0.6\;& 0.65  \\ \hline \hline
${\theta}$\;\;& 0.725 &0.700 & 0.670\;\; &0.639 &\; 0.612 \;\;&0.586 & 0.563 \;&0.541 & 0.521 &0.503 & 0.486 & 0.469   \\  \hline
${\kappa}$\;\;&0.440 &0.472 & 0.496\;\; &0.515 &\; 0.532\;\;&0.547 & 0.560\;\;&0. 572 &\; 0.583\;\;&0.594 &\; 0.604\;\; &\;\;0.613 \;\;    \\ \hline \\
 \hline
$a (\mu m)$\; &0.70&0.75 & 0.8 & 0.85 & 0.9& 0.95&  1 &1.2&1.4 &1.6 & 1.8 & 2  \\ \hline \hline
${\theta}$ \;\; &0.454 &0.440 &0.426&0.413 &0.401&0.389&0.378 &0.339&0.307&0.279& 0.256&0.236   \\  \hline
${\kappa}\;\;$&0.622&\;0.631 \;&0.639 &0.647&0.655&0.662&\; 0.670\;\;&0.696&0.712&0.739&0.758&0.774\\ \hline  \hline
\end{tabular}
\caption{Values of the coefficients ${\theta}$ and ${\kappa}$ for Au at room temperature (plasma prescription).}
\end{table*}
\end{center}
\label{tab.4}
\begin{center}
\begin{table*}
%\begin{tabular}{l l l l l l l l l l l | l  l} \hline
\begin{tabular}{ccccccccccccc} \hline
$a (\mu m)$\;\; &0.10& 0.15 & 0.2  \;\;& 0.25\;\;\;& 0.3\;\; &0.35 \;\;& 0.4 \;\; & 0.45 & 0.5\;\;& 0.55 & 0.6\;\; & 0.65  \\ \hline \hline
${\tilde \theta}$\;\;& 0.463  &\;0.477 &\; 0.475\;\; &0.467 & \;0.458 \;\;&0.447 & 0.437 \;&0.427 & 0.417 &0.407 & 0.398&0.389    \\  \hline
${\tilde \kappa}$\;\;&0.244 &\;0.269 &\; 0.289\;\; & 0.306 & 0.319 \;&0.332 & 0.343\;\;&0.352  &\; 0.363\;&0.372 & \;\;0.380\;\; &\;0.389 \;\;    \\ \hline \\
 \hline
$a (\mu m)$\; &0.70&0.75 & 0.8 & 0.85 & 0.9& 0.95&  1 &1.2&1.4 &1.6 & 1.8 & 2  \\ \hline \hline
${\tilde \theta}$ \;\; &0.380 &0.371  &0.363 &0.355 &0.348&0.340&0.333 &0.306&0.282&0.261& 0.242 &0.225  \\  \hline
${\tilde \kappa}\;\;$&0.397&\;0.406 \;&0.413 &0.421&0.430&0.437&\; 0.444\;\;&0.474&0.502&0.529&0.554&0.578\\ \hline  \hline
\end{tabular}
\caption{Values of the coefficients ${\tilde \theta}$ and $\tilde \kappa$ for Au at room temperature (plasma prescription).}
\end{table*}
\end{center}
\end{widetext}

Now that we have addressed the contribution of the positive Matsubara modes, we can turn our attention to the zero-frequency contributions. These contributions shall be dealt with in the next two sections, first within the Drude prescription and then within the plasma prescription.

\section{The zero-frequency contribution}

\subsection{Drude prescription: grounded vs isolated conductors}

Within the Drude prescription, a conductor is transparent to static magnetic fields. This implies that  TE modes do not contribute to the scattering formula for zero frequency. On the contrary, zero-frequency TM modes, which represent electrostatic fields,  are screened out by a Drude conductor and therefore  they do contribute with full power to the $n=0$ term of the scattering formula. 

\begin{figure}
\includegraphics[width=.9\columnwidth]{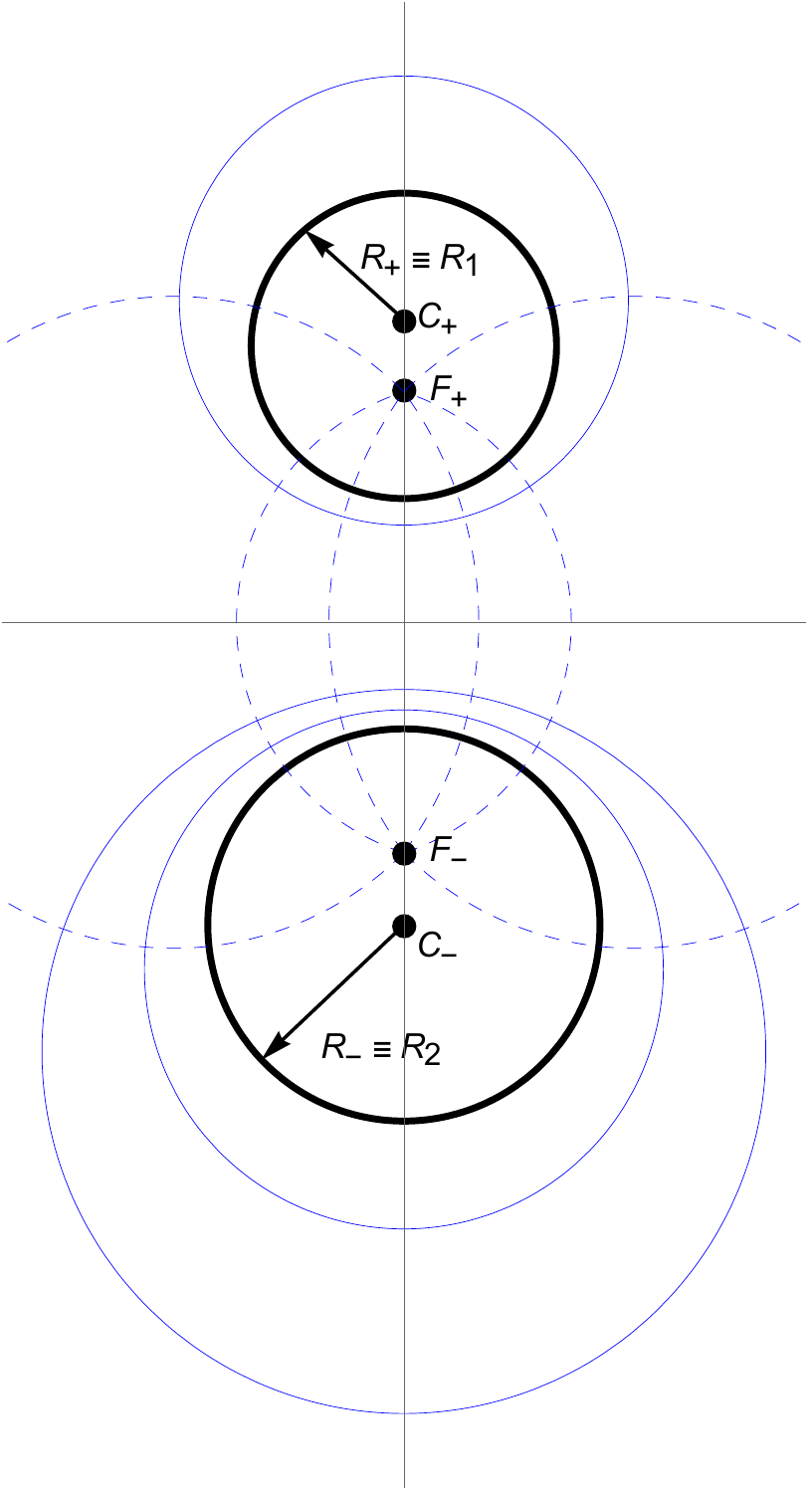}% Here is how to import EPS art
\caption{\label{biscoo}  The two-sphere system in bispherical coordinates. The thin solid and dashed lines are curves of constant bispherical coordinates $\mu$ and $\eta$ respectively. The sphere of radius $R_1$   has equation $\mu=\mu_+$, with $\mu_+>0$, while the sphere of radius $R_2$   has equation $\mu=\mu_-$, with $\mu_-<0$. Bispherical partial waves have their origins at the foci $F_{\pm}$, while standard spherical multipoles have their origins at the sphere centers $C_{\pm}$.}
\end{figure}

To compute the contribution of zero-frequency TM modes,   we take advantage of the fact that the Laplace equation obeyed by the electrostatic potential $\phi$   is separable in bispherical coordinates \cite{morse}. We recall that bispherical coordinates $(\mu, \eta, \phi)$  are defined by $(x,y,z)=b (\sin \eta \cos \phi, \sin \eta \sin \phi, \sinh \mu)/(\cosh \mu-\cos \eta)$, where $\pm b$ represent the $z$-coordinates of  the foci $F_{\pm } \equiv (0,0 \pm b)$ of the spheres of equation $\mu=\mu_{\pm}$, where $\mu_+>0$ and $\mu_-<0$. The spheres have radii $R_{\pm}=b/|\sinh \mu_{\pm}|$, and their centers are placed at the points $C_{\pm}$ of the $z$-axis of coordinates $z_{\pm}= b \coth \mu_{\pm}$  (see Fig.\ref{biscoo}). Below, we shall use the notation $R_1\equiv R_+$ and $R_2 \equiv R_-$. The regular and   outgoing eigenfunctions of Laplace equation in bispherical coordinates are \cite{morse}:
\begin{eqnarray}
\phi_{lm}^{(\pm)}&=&\sqrt{2( \cosh \mu-\cos \eta)} \,\nonumber \\
&\times &  \exp[\pm (l+1/2)\mu]\,P^{|m|}_l (\cos \eta)\,e^{i m \phi}\;,\label{bisph}
\end{eqnarray}
where $l \ge 0$, and $m=-l, \cdots, l$. Relative to the sphere $R_+$ ($R_-$)  outgoing and regular eigenfunctions correspond, respectively  to the upper (lower) and lower (upper) sign in the exponential. We thus see that in the bispherical coordinate system,  a {\it single} set of partial waves serves as a basis of scattering states for {\it both} spheres. This implies at once that in bispherical coordinates the translation operator ${\hat U}$ is just the identity, which in turn implies that the ${\hat N}$ operator reduces to the product of the T-operators of the spheres:
\be
{\hat N}={\hat T}^{(-)}{\hat T}^{(+)}\;,\label{N1bisp}
\ee 
where here and below we do not display from brevity the dependence of all matrices on the azimuthal number $m$.
 It remains to compute the matrix elements of the T-operators $T^{(\pm)}_{l,l'}=\langle l,m,\pm | {\hat T}^{(\pm)}| l',m,\mp \rangle$ of the two spheres in the bispherical basis, where we set $| l,m,\pm \rangle \equiv \phi_{lm}^{(\pm)} $. 

The expression of the T-matrix of a conducting sphere depends on whether the sphere is grounded or not. If the spheres are grounded (or what is the same, connected to a charge reservoir) 
the fluctuations of the electrostatic potential $\phi$ satisfy Dirichlet (D) boundary conditions (bc) on the surfaces of the spheres, and then it is a simple matter to check that the matrix elements of the T-operators are:
\be
T^{(\pm)}_{l,l'}=\delta_{l,l'}\,Z_{\pm}^{2 l+1}\;,\;l,l' \ge |m|\;\;\;{\rm (D\;bc)}\;\label{DTmat}
\ee
where $Z_{\pm}=\exp[\mp \mu_{\pm}]$.  Since the  T-matrices of the spheres are diagonal,  evaluation of  the scattering formula is straightforward, yielding the Casimir energy:
\be
{\cal F}_{n=0}^{(\rm D)}=\frac{k_B T}{2} \sum_{l=0}^{\infty} (2 l +1) \ln[1-Z^{2 l+1}]\;,\label{diren}
\ee
where $Z=Z_+ Z_-$, and we introduced  the superscript $(\rm D)$ to remind us that this is for  D bc. The parameter $Z$ has the following expression in terms of $R_1, R_2$ and $a$:
\be
Z=[1+x+x^2 u/2+\sqrt{(x+x^2 u/2)\,(2+x+x^2 u/2)}]^{-1}\;,\label{Zdef}
\ee
where $x=a/\tilde{R}$.
 The energy (\ref{diren}) was derived in \cite{bimonteex1} and it was used in \cite{bimonte2sphere} to study the two sphere problem with grounded spheres. 

At this point we turn our attention to the case of two {\it isolated} spheres. This is the configuration usually understood in the Casimir literature by the expression Drude bc. When evaluated in the standard spherical basis with origins placed in the spheres centers, the matrix elements of the $T$-matrix coincide with the familiar Mie coefficients. What we want to do here is to re-express the zero-frequency Mie coefficients in the bispherical coordinate system.    For two isolated spheres, the fluctuations of the electrostatic potential are subjected to the constraint that the fluctuation of the total charge on either sphere is zero. This seemingly innocent condition complicates considerably the task of computing the Casimir energy. The partial waves in Eq. (\ref{bisph}) are not fit to Drude bc, because for $m=0$ they include an undesired monopole contribution. This can be seen by expanding $\phi_{lm}^{(\pm)}$ in spherical multipoles with origins at the foci $F_{\pm}$:
\be
\phi_{l'm}^{(\pm)}=\sum_{l =|m|}^{l'} \frac{(l'+|m|)!}{(l+|m|)!(l'-l)!}\left(\frac{2 b}{{\hat r}_{\pm}}\right)^{l+1}P^{|m|}_l (\cos {\hat \theta}_{\pm})\,e^{i m \phi}\,,\label{bispsp}
\ee
where ${\hat r}_{\pm}$ is the radial distance from $F_{\pm}$, and the angles ${\hat \theta}_{\pm}$ are defined such that  ${\hat \theta}_{\pm}=0$ for $\eta=0$. From this expansion,
we see that for $m \neq 0$ the partial waves $\phi_{l'm}^{(\pm)}$ are free from monopole contributions, and therefore they automatically respect the constraint of total zero charge. This implies at once that the contribution to the Drude Casimir energy of the modes with $m \neq 0$ is identical to the corresponding contributions in the D case. This in turn permits to write the Drude Casimir energy  ${\cal F}^{(\rm Dr)}_{n=0}$ as:
\be
{\cal F}^{(\rm Dr)}_{n=0}={\cal F}^{(\rm Dr)}_{n=0}\vert_{m=0}+\frac{k_B T}{2} \sum_{l=1}^{\infty} 2 l  \ln[1-Z^{2 l+1}]\;,\label{splitDr}
\ee   
where ${\cal F}^{(\rm Dr)}_{n=0}\vert_{m=0}$ represents the contribution of the modes with $m=0$. We denote by ${\hat N}_0$, ${\hat T}^{(\pm)}_0$  the restrictions of the Drude operators ${\hat N}$ and ${\hat T}^{(\pm)}$, respectively, to the $m=0$ subspace, and then ${\cal F}^{(\rm Dr)}_{n=0}\vert_{m=0}$ is:
\be
{\cal F}^{(\rm Dr)}_{n=0}\vert_{m=0}=\frac{k_B T}{2}\,{\rm \ln \det} [1-{\hat N}_0 ]\;,
\ee 
where
\be
{\hat N}_0={\hat T}_0^{(-)}{\hat T}_0^{(+)}\;,\label{Nbisp}
\ee
and it is understood that the determinant is restricted to the $l,l'$ space. To remove the undesired monopole term ${\hat r}_{\pm}^{-1}$ from $\phi_{l0}^{(\pm)}$ ,  for $m=0$ we consider a new set of partial waves ${\bar \phi}_{l}^{(\pm)}$ defined as:
\be
{\bar \phi}_{l}^{(\pm)} \equiv \phi_{l 0}^{(\pm)}-\phi_{0 0}^{(\pm)}\;,\;\;\;l \ge 1.
\ee
To determine the  matrix elements of the Drude T-operator $(T^{(\pm)}_0)_{l,l'}$ in the basis of the  ${\bar \phi}_{l }^{(\pm)}$, we proceed as follows. Consider  scattering  the outgoing wave  ${\bar \phi}_{l' }^{(\pm)}$, originating from the sphere $R_{\pm}$, by the sphere $R_{\mp}$. For the scattered field one can make the ansatz: 
\be
\phi^{(\rm scat|\mp)}_{l'}\vert_{\rm Dr}=\phi^{(\rm scat|\mp)}_{l'}\vert_{\rm D}- \frac{2 \,b}{\;\;r_{\mp}} k_{l'}^{(\mp)}\;,\label{scatDr}
\ee 
where $\phi^{(\rm scat|\pm)}_{l'}\vert_{\rm D}$ denotes the scattered field for D bc, and  $2 b /r_{\pm}$ is the potential of a charge placed at the center of the sphere $R_{\pm}$ ($r_{\pm}$ is the radial distance from the center of the sphere $R_{\pm}$). The D scattered field $\phi^{(\rm scat|\mp)}_{l'}\vert_{\rm D}$ is easily computed with the help of Eq. (\ref{DTmat}).  The coefficient $k_{l'}^{(\pm)}$ is  determined such as to cancel the monopole terms present in $\phi^{(\rm scat|\pm)}_{l',0}\vert_{\rm D}$. A straightforward computation yields:
\be
k_{l'}^{(\pm)}=e^{\mp\mu_{\pm}}(e^{\mp2\, l' \mu_{\pm}}-1)\;.
\ee
At this point, all we have to do is to expand the scattered field on the rhs of Eq. (\ref{scatDr}) in the basis of the regular waves ${\bar \phi}^{(\pm)}_{l}$ of the sphere $R_{\mp}$. The latter task is easily accomplished with the help of the following translation formula:
\be
\frac{R_{\pm}}{r_{\pm}}=\sum_{l=0}^{\infty}e^{-l |\mu_{\pm}|} \left(\frac{R_{\pm}}{{\hat r}_{\pm}}\right)^{l+1}P_l (\cos {\hat \theta}_{\pm})\,,
\ee
together with the inverse of Eq. (\ref{bispsp}) for $m=0$
\be
 \left(\frac{2 b}{{\hat r}_{\pm}} \right)^{l'+1 } P_{l'} (\cos {\hat \theta}_{\pm})\;=\sum_{l=0}^{l'} \frac{ (-1)^{l+l'}  l'!}{(l'-l)!\, l!} \phi^{(\pm)}_{l,0}\;.
\ee
The final result is
\be
(T^{(\pm)}_0)_{l,l'}= \left[\delta_{ll'}+(1-Z_{\pm}^2)(1-Z_{\pm}^{2 l'}) \right]\,Z_{\pm}^{2 l+1}\;.\label{TDr}
\ee
As we see,  the matrices $T^{(\pm)}_0$ of the spheres are non-diagonal, which renders an exact evaluation of the scattering formula impossible. A remarkable exception occurs though in the sphere-plate geometry. The  configuration of a sphere of radius $R$ facing a plane corresponds, for example,  to the choice  $\mu_-=0$ (which is the equation of the $z=0$ plane) and $R_+ \equiv R$. From Eq. (\ref{TDr}) we see that for  $\mu_-=0$  the matrix $T^{(-)}_0$ becomes the identity. Equation  (\ref{Nbisp}) then shows that the $N_0$-matrix of the sphere-plate system in bispherical coordinates coincides with  the matrix $T^{(+)}_0$ of the sphere. At this point, one makes the crucial observation that the energy ${\cal F}^{(\rm Dr)}_{n=0}\vert_{m=0}$ actually depends only on the equivalence class $[[N_0]]$  formed by the matrices that represent the operator ${\hat N_0}$, where any two elements within the equivalence class differ by a similarity transformation by some non-singular matrix $M$. It is easy to verify that by  an appropriate choice of $M$ it is possible to express the equivalence class of  ${\hat N}_0$   as:
\be
{\hat N}_0=[[ \left[\delta_{ll'}+(1-Z^2)(1-Z^{2 l'}) \right]\,Z^{2 l'+1}]]\;,
\ee
where we used also the relation $Z=Z_+$ that holds in the sphere-plate case.
The peculiar feature of  the matrix on the rhs of the above equation is that it consists of a diagonal matrix plus a matrix whose elements depend only on the column index $l'$. In \cite{bimonteex1} it is shown that such a structure allows for a direct computation of $\det (1-{\hat N}_0)$,  which together with Eq. (\ref{splitDr}) leads to the following remarkable formula for ${\cal F}^{(\rm Dr)}_{n=0}$ in the sphere-plate case:
$$
{\cal F}_{n=0}^{(\rm Dr)}\vert_{\rm sp-pl}=\frac{k_B T}{2} \left\{\sum_{l=1}^{\infty} (2 l +1) \ln[1-Z^{2 l+1}]\;\right.
$$
\be
\left.+\ln \left[1-(1-Z^2) \sum_{l=1}^{\infty} Z^{2 l+1} \frac{1-Z^{2 l}}{1-Z^{2 l+1}} \right] \right\}\;.\label{enerDrsp}
\ee
Unfortunately, for two spheres of finite radii $R_{\pm}$, no such tricks seem to exist, and therefore no direct computation $\det (1-{\hat N}_0)$  is possible. The latter determinant has to be computed numerically in this case, and here too bispherical coordinates display their power, since the number of bispherical multipoles that are needed to achieve convergence scales just like $\sqrt{{\tilde R}/a}$, instead of ${\tilde R}/a$ which is the scaling  law in the standard representation with spherical multipoles.   As a result of this improvement, for aspect ratios ${\tilde R}/a$ as large as $10^4$  it is possible to compute $\det (1-{\hat N}_0)$ very accurately with just 100 bispherical multipoles or so, an easy task for a laptop.  

It is interesting to compare the forces  ${F}_{n=0}^{(D)}$ and $F_{n=0}^{(\rm Dr)}$, to see what is the effect of grounding the spheres. The first thing to notice is that  ${F}_{n=0}^{(D)}$ and $F_{n=0}^{(\rm Dr)}$ have the same asymptotic expansion in the limit $a/{\tilde R} \rightarrow 0$, up to order ${a/\tilde{R}}$:
\be
 {F}_{n=0}^{(\rm D)}\simeq  {F}_{n=0}^{(\rm Dr)} =-k_B T\frac{ \zeta(3) {\tilde R}}{8\, a^2} \left(1+\frac{1}{6 \zeta(3)} \frac{a}{\tilde R} +o(a/{\tilde R}) \right)\;,\label{DEnzero}
\ee
where   $\zeta(x)$ is Riemann zeta function.
The leading term coincides with the PFA Eq. (\ref{PFA}), since   ${\cal F}^{(\rm pp)}_{n=0}\vert_{\rm Dr}={\cal F}^{(\rm pp)}_{n=0}\vert_{\rm D}=-k_B T \zeta(3)/(16 \pi a^2)$.  The leading correction beyond PFA, i.e. the term proportional to $a/{\tilde R}$  on the rhs of the above equation,   is consistent with the DE (see \cite{bimonte2sphere}) and interestingly it is independent of the parameter $u$. 
 
While coinciding in the regime of small separations, ${F}_{n=0}^{(\rm D)}$ and $ {F}_{n=0}^{(\rm Dr)}$ do have drastically different behaviors in the opposite limit of large separations $a/{\tilde R} \rightarrow \infty$. In the language of bispherical coordinates, the large distance limit corresponds to taking 
$\mu_{\pm} \rightarrow \pm \infty$, i.e. $Z_{\pm} \rightarrow 0$. In this limit, one easily finds:
\be
{F}_{n=0}^{(\rm D)}=-k_B T \frac{R_1 R_2}{a^3}\left[1+O\left(\frac{R_1+R_2}{a} \right)\right] \;,
\ee
\be
{F}_{n=0}^{(\rm Dr)}=- 18\, k_B T \frac{R_1^3 R_2^3}{a^7}\left[1+O\left(\frac{R_1+R_2}{a} \right)\right]\;. 
\ee

We recall that in all Casimir experiments,   the conductors  are always connected to charge reservoirs. This configuration is used  to get rid of possible stray charges that could be present on the surfaces, and to compensate by a suitably adjusted bias potential for unavoidable differences between the work functions of the conductors and/or potential patches \cite{deccapat}. The  correct model for the   TM zero-frequency mode of the experimental setups  is therefore  represented by the D bc, rather than the commonly considered Drude bc. Since according to Eq. (\ref{DEnzero}) grounded and isolated spheres have the same energy  up-to first-order beyond PFA, it is clear that the differences between the two models show up only at the level of   second-order corrections beyond PFA,  whih are exceedingly small for typical experimental values of the aspect ratio. This explains why experimentally, there has be no need so far to make a distinction between the two cases. In the next section the deviations from PFA of grounded and isolated spheres shall be computed numerically.

\subsection{Perfect conductor} 

We consider now the zero-frequency contribution for a perfect conductor (PC).  We study this model because its zero-frquency term can be easily computed numerically in bispherical coordinates, and more importantly because for separations larger than a few times the plasma length, the deviation from PFA for a PC  provides a very good approximation  to the deviation from PFA for the zero-frequency term of the plasma model, to be studied in the next subsection.  

The difference between  the Drude model, that was studied in the previous subsection, and  a PC  consists in the fact that the latter expels not only electrostatic fields but also  static magnetic fields (perfect Meissner effect). As a result of this circumstance, in addition to the contribution of the TM modes, which is identical to the classical term of the Drude model, the PC classical term receives a contribution also from the TE modes.  For a PC the latter  is identical to the classical term of a scalar field subjected to Neumann (N) bc \cite{antoine}. 

The classical Casimir interaction of PC sphere and plate was investigated numerically in \cite{antoine}, using a large simulation of the scattering formula in the standard spherical multipole basis.
The same problem was studied in \cite{bimonteHT} using bispherical coordinates. What renders interesting the PC classical Casimir interaction is that the DE fails for a classical N field, since the associated perturbative kernels  lack the analiticity properties for small in-plane momenta  that are necessary for existence of the DE. Indeed, in the sphere-plate geometry, the leading correction beyond PFA for a classical N field has a $\log^2 (R/a)$ dependence \cite{antoine,bimonteHT}, instead of the $a/R$ dependence that is characteristic of situations in which the DE is valid (see \cite{fosco1,bimonte1,fosco2,fosco3,bimonte2}).

Here we use the bispherical coordinates to study the classical PC two-sphere problem. As we said earlier, the PC TM classical contribution is identical to the Drude case, and so it remains to compute the TE contribution, i.e. the N classical term.   Similar to the Drude model studied in the previous section, Eq. (\ref{N1bisp}) holds also to the N problem, and therefore all we need to determine ${\hat N}$ are the bispherical matrix elements $T^{\rm (N|\pm)}_{l,l'}$  of the T-operators the spheres, where the superscript stands for Neumann. It is convenient to express $T^{\rm (N|\pm)}_{l,l'}$  as a perturbation of the D case, and thus we decompose $T^{\rm (N|\pm)}_{l,l'}$ in the form:
\be
T^{\rm (N|\pm)}_{l,l'}=-Z_{\pm}^{ l+1/2} Z_{\pm}^{ l'+1/2}\left[\delta_{l,l'}+\delta T^{(\pm)}_{l,l'} \right]\:,\;\;\;l,l' \ge |m|.\label{factN}
\ee
The matrix $T^{\rm (N|\pm)}_{l,l'}$ is computed as follows. One considers again scattering  the outgoing wave   $\phi_{l'm}^{(\pm)}$, originating from the sphere $R_{\pm}$, by the sphere $R_{\mp}$. By definition of the T-matrix, the total field $\phi_{l'm}^{(\rm tot|\mp)}$  outside the sphere $R_{\mp}$  is the sum of the incoming field plus the scattered field:
\be
\phi_{l'm}^{(\rm tot|\mp)}= \phi_{l'm}^{(\pm)}- \sum_{l} \phi_{lm}^{(\mp)} T^{\rm (N|\mp)}_{l,l'}\;.\label{totN}
\ee
 The T-matrix for N bc is then determined by demanding  that the total field satisfies the condition:
\be
\partial_{\mu } \phi_{l'm}^{(\rm tot|\mp)}\vert_{\mu=\mu_{\mp}}=0\;
\ee
on the surface of $R_{\mp}$. After substituting Eq. (\ref{totN}) into the above equation, and then making use of the the expression of the T-matrix  in Eq. (\ref{factN}), it is not hard to prove the following {\it exact} equation that has to be satisfied by the matrix $\delta T^{(\pm)}$:
\be
B^{(\pm)}  \;\delta T^{(\pm)}= \mp 2 \sinh \mu_{\pm} \;1\;,\label{linT}
\ee
where $B^{(\pm)}  $ is the matrix of elements
$$
B^{(\pm)}_{ll'}=[(2l+1) \cosh \mu_{\pm}\pm\sinh \mu_{\pm}] \delta_{ll'}
$$
\be
-(l-|m|)\delta_{l,l'+1}-(l'+|m|) \delta_{l+1,l'}\;.
\ee
The linear equations satisfied by $\delta T^{(\pm)}$ cannot be solved analytically, but can be easily solved numerically. We note that according to Eqs. (\ref{scattF}) and (\ref{scattF'}) the computation of $F$ and $F'$ involve the first and second derivatives of the T-matrix with respect to the separation, which in turn involve $\mu$-derivatives of  $\delta T^{(\pm)}$.
The linear equation satisfied by the $\mu$-derivatives of  $\delta T^{(\pm)}$  are easily found by taking the $\mu$-derivatives of both sides of Eq. (\ref{linT}). 

The N classical force and force-gradient can be computed numerically in bispherical coordinates very quickly because also for N bc, the scattering formula converges in these coordinates for a multipole order that scales like $\sqrt{{\tilde R}/a}$. As a demonstration of the improved rate of convergence in bispherical coordinates, we recall  that the values of the N sphere-plate classical energy that were obtained in \cite{antoine} by including up to 5000 spherical multipole for  aspect ratio $R/a \simeq 10^3$, can be reproduced by including just 100 bispherical multipoles \cite{bimonteHT}.

\subsection{Plasma model}

We saw earlier that both the Drude and the PC models can be addressed efficiently in bisperical coordinates. Unfortunately, this is not possible with the plasma model. Within this prescription, the classical Casimir interaction receives a contribution from both TM modes and TE modes. The TM contribution is identical to the Drude case,  grounded or isolated depending on the experimental configuration, and therefore it can be handled in bispherical coordinates. The troublesome contribution is the TE one. We said earlier that TE modes describe classical fluctuations of the  magnetic field. The vector potential of the latter satisfies inside the spheres the vector Helmoltz equation, which is not separable in bispherical coordinates. Therefore for the TE mode there is no alternative to using the standard spherical multipole basis. In the latter basis,   and for zero frequency, the plasma model $N$-operator has the matrix representation \cite{pablo,teo2}:
\be
N=A^{(2)} A^{(1)}\;,
\ee
where the matrices $A^{(i)}$ are:
\be
A_{l,l'}^{(i)}=\frac{(l+l')!}{(l+m)!(l'-m)!} \left(\frac{R_i}{L}\right)^{2 l'+1} r^{(i)}_{l'}\;,
\ee
where $  l,l' \ge |m|$,  $L=R_1+R_2+a$ is the center-to-center distance,  and
\be
r^{(i)}_l= \frac{l}{l+1}\,\frac{I_{l+3/2}(\omega_p^{(i)} R_i/c)}{I_{l-1/2}(\omega_p^{(i)} R_i/c)}\;.
\ee
In the above equation, $ I_{p}(z)$ are the modified Bessel functions of the first kind, while $\omega_p^{(i)}$ is the plasma frequency of the sphere of radius $R_i$. The PC model is recovered from the plasma model in the limit   $\omega_p^{(i)} R_i/c \rightarrow \infty$, in which  $r^{(i)}_l$ approaches the expression for N bc: $r^{(i)}_l \rightarrow l/(l+1)$.

For the purpose of numerical computations, it is convenient to take advantage of the freedom to conjugate $N$ by a non-singular matrix $M$, to convert $N$ to the following symmetric form:
\be
[[N]]=A A^{\dagger}\;, 
\ee 
where
\begin{widetext}
\be
A_{l,l'}=\left(\frac{R_1}{L}\right)^{l+1/2}\sqrt{r_l^{(1)}}\frac{(l+l')!}{\sqrt{(l+m)!(l-m)!(l'+m)!(l'-m)!}}\left(\frac{R_2}{L}\right)^{l'+1/2}\sqrt{r_{l'}^{(2)}}\;, \;\;\;l,l' \ge |m|\;.
\ee
\end{widetext}
The improved numerical stability ensured by a symmetrized form of the $N$-matrix   is extensively discussed in \cite{gert}. 

When the scattering formula is computed numerically, it is obviously necessary to truncate the multipole indices $l$, $l'$ and $m$ below some maximum orders $l_{\rm max}$, $l'_{\rm max}$ and $m_{\rm max}$, respectively. In order to gain a precise feeling of how far one needs to go with the multipole order, it is useful to examine more closely the matrix $A$. One is interested here in investigating how the matrix elements of $A$ behave for large aspect ratios ${\tilde R}/a \gg 1$. A simple estimate of the magnitude of the elements of the matrix $A$ can be obtained by taking  their logarithms, and then using Stirling formula for the factorials. When  doing that, the multipole indices $l$, $l'$ and $m$  can all be  treated as continuous variables. It is useful to set $l=l_0+\tilde{l} $ and $l'=l_0 R_2/R_1 +{\tilde l}'$, and then one finds:
\begin{widetext}
\be
A_{l,l'} \simeq  \sqrt{r_l^{(1)} r_l^{(2)} } \sqrt{\frac{R_1}{2 \pi (R_1+R_2)l_0}}\;\exp \left[-\frac{a}{R_1} l_0-\frac{({\tilde l R_2-{\tilde l}' R_1})^2}{2 l_0 R_2(R_1+R_2)} -\frac{R_1+R_2}{2 R_2 l_0} m^2\right]\;.\label{Astir}
\ee   
\end{widetext}
The above expression provides us with a wealth of information about the matrix A. First of all, it shows that the multipole orders scale according to the following laws:   $l_{\rm max} \approx R_1/a$, $ l'_{\rm max} \approx R_2/a$   and $m_{\rm max} \approx \sqrt{{\tilde  R}/a}$ (in practice, in our numerical simulations we do observe convergence for $l_{\rm max}-|m| = 6 R_1/a \equiv N_1$, $ l'_{\rm max}-|m| =6 R_2/a \equiv N_2$   and $m_{\rm max} = \sqrt{6{\tilde  R}/a}$). Second, Eq. (\ref{Astir}) shows that the rectangular matrix $A$ of size $N_1 \times N_2$  is indeed dominated by the elements comprised within the oblique strip $\Sigma$ of "horizontal" half-width $\Delta_1 \simeq R_1/\sqrt{{\tilde R} a}$ and "vertical" half-width $\Delta_2 \simeq R_2/\sqrt{{\tilde R} a}$ around the "main diagonal" $l \simeq l' R_1/R_2$ of the  matrix $A$. A good deal of computer time can indeed be saved by just storing the elements of $A$ contained within the strip $\Sigma$. Despite this, one runs into trouble for aspect ratios ${\tilde R}/a$ exceeding 100 or so.  Large aspect ratios can  however be handled by  a simple decimation procedure, that allows to compress the size of the matrix $A$, as shown in the next section.

\section{A Decimation scheme for the scattering formula}

For large aspect ratios the computation of the classical term for the plasma model becomes prohibitively time consuming. For example, the precise computation of the force for two spheres with a radius of 50 micron at a minimum separation of 100 nm requires going up to multipole index $l_{\rm max} \simeq$ 3000, which is  an impossible task for  an ordinary laptop.  To handle large aspect ratios, we developed a very simple but effective  decimation procedure, that permits to compress the size of the matrices that need to be computed, without jeopardising the precision of the computation. The decimation scheme works as follows.

One notes first that the scattering formulae for the Casimir energy,  force and force gradient can be all expressed as sums of traces of products of the matrix $A$ and its derivatives. Consider for example the  Casimir energy. By exploiting the identity $\ln \det B = {\rm Tr} \ln B$, that holds for any positive matrix, one can write the classical term (which is what we are after here, but the same thing can be done of course for all Matsubara modes) of the scattering formula Eq. (\ref{scattF}) as:
\be
{\cal F}_{n=0}=\frac{k_B T}{2} {\rm Tr}\ln [1- (A A^{\dagger})]=-\frac{k_B T}{2} \sum_{k=1}^{\infty} \frac{1}{k}{\rm Tr}\left(A A^{\dagger} \right)^k\;. \label{sersca}
\ee 
Now comes the crucial observation: according to Eq. (\ref{Astir}),  the matrix elements of $A$  do not change appreciably if its indices $l$ and $l'$ are increased by amounts $k$ and $k'$ whose magnitudes are small compared to $\Delta_1$ and $\Delta_2$, respectively, i.e. $A_{l,l'} \simeq A_{l+k,l'+k'}$ for $|k| \ll \Delta_1$, $|k'| \ll \Delta_2$ (where for brevity we suppressed the index $m$).  Consider then two submultiples $p_1$ and $p_2$ of $N_1$ and $N_2$, respectively, such that $1<p_1\ll \Delta_1$ and $1 < p_2 \ll \Delta_2$, and let  $N_i/p_i \equiv n_i\;,\;i=1,2$. Now imagine subdividing the matrix $A$ into $n_1 \times n_2$  blocks ${\cal B}_{j_1,j_2}\;,\;j_i=1,\dots n_i$, each of size $p_1 \times p_2$. The blocks are numbered such that
within ${\cal B}_{j_1,j_2}$ the indices $l,l'$ span the range $p_1(j_1-1) \le l-|m| < p_1 j_1$ and  $p_2(j_2-1) \le l'-|m| < p_2 j_2$.  Bearing in mind the previous observation about the slow variation of $A_{ll'}$, the matrix elements of $A$ can be considered as practically {\it constant} within each block  ${\cal B}_{j_1,j_2}$. It is therefore legitimate to expect that a small error would be made in the evaluation of the scattering formula if  the exact matrix $A$ was replaced by the  $N_1 \times N_2$ matrix ${\bar A}$ whose elements are {\it constant} within each block  ${\cal B}_{j_1,j_2}$, and coincide with a representative element $a_{j_1,j_2}$ chosen among the $p_1 \times p_2$ matrix elements of $A$ within  ${\cal B}_{j_1,j_2}$. Consider at this point the $n_1 \times n_2$ matrix $A_{\rm dec}$ formed by the chosen representatives $a_{j_1,j_2}$: $(A_{\rm dec})_{j_1,j_2}=a_{j_1,j_2}$. It is not hard to convince oneself that  for each term in the series in Eq. (\ref{sersca}) the approximate identity holds:
\be
{\rm Tr}\left(A A^{\dagger} \right)^k \simeq{\rm Tr}\left({\bar A}{\bar A}^{\dagger} \right)^k= (p_1 p_2)^k{\rm Tr}\left(A_{\rm dec} A_{\rm dec}^{\dagger} \right)^k\;.
\ee
This identity implies that the value of  ${\cal F}_{n=0}$ should be approximately invariant under the substitution of the $N_1 \times N_2$ matrix $A$ by the following decimated and {\it rescaled} $n_1 \times n_2$ matrix ${\tilde A}$:
\be
A \rightarrow {\tilde A}=\sqrt{p_1 p_2}A_{\rm dec}\;,\label{Adec}
\ee
\be
{\cal F}_{n=0} \simeq \frac{k_B T}{2} {\rm Tr}\ln [1-   {\tilde A}\,{\tilde A}^{\dagger}]\;.\label{decim}
\ee
The substitution of the matrix $A$ by the matrix $\tilde A$   works as well for the Casimir force and force gradient.
We tested Eqs. (\ref{Adec}) and (\ref{decim}) in the N case, where the Casimir force can be computed to high precision using bispherical coordinates up to large aspect ratios.   It was  found that the decimation procedure   nicely reproduces the bispherical values of the force and force gradient,  for values of the aspect ratio ${\tilde R}/a$ around 1000, using blocks of size around $10 \times 10$. We made sure that the decimated $N$-matrix $1-   {\tilde A}\,{\tilde A}^{\dagger}$ is still positive for blocks of this size, and we found that the best results were obtained by picking for the  representatives $a_{j_1,j_2}$  the elements of $A$ in the upper left corners of the blocks  ${\cal B}_{j_1,j_2}$. To give the reader a quantitative sense of the effectiveness of the decimation method, the following example may suffice.  For two spheres of radii $R_1=R_2=50$ micron at a separation $a=100$ nm, using $7 \times 7$ blocks, the percent error $\delta$ in the force, engendered by the decimation procedure,  is  $\delta=$0.1 \% for the $m=0$ modes. The error becomes smaller and smaller for larger and larger values of $m$. For example, for $m=1,2$ and $3$ the errors are $\delta=$0.05 \%, 0.009 \%  and 0.00025 \%, respectively. This is good enough to obtain reliable estimates of the small deviations from PFA we are after.

\section{Numerical computations}

In this section, we display the results of our numerical computations.  The latter are better presented by introducing  the following quantities $\beta$ and $\tilde \beta$ that measure the deviations of the computed forces from the PFA:
\be
\beta=\frac{\tilde R}{a}\left(\frac{F}{F_{\rm PFA}}-1 \right)\;,
\ee   
and
\be
{\tilde \beta}=\frac{\tilde R}{a}\left(\frac{F'}{F'_{\rm PFA}}-1 \right)\;.
\ee   

\begin{figure}
\includegraphics[width=.9\columnwidth]{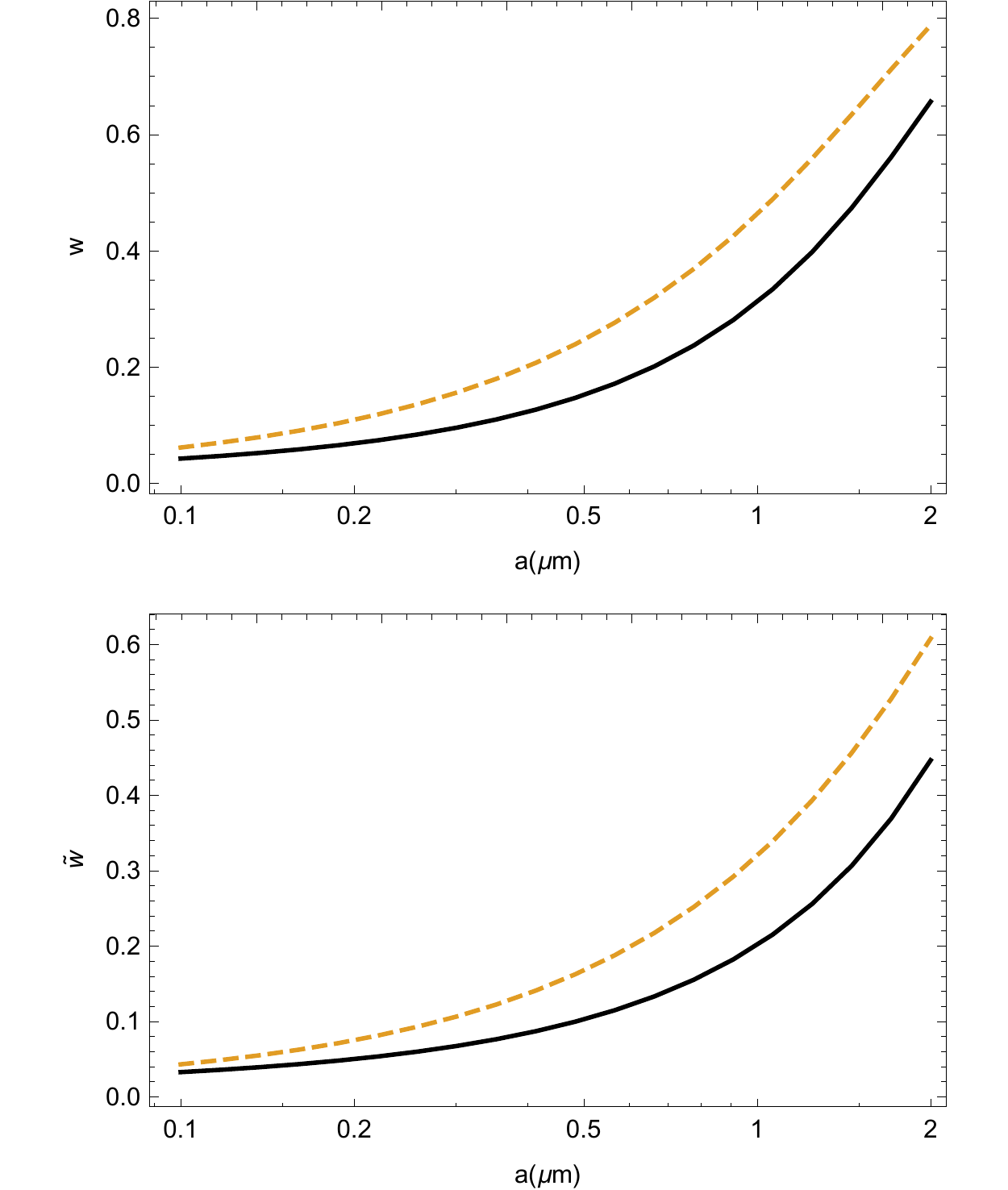}% Here is how to import EPS art
\caption{\label{wplot}  The upper panel shows the ratio between the zero-frequency contribution to the Casimir force and the total Casimir force, while the lower planel displays the analogous ratio for the force gradient. All forces are computed for gold at room temperature, using the PFA. Solid lines are for the Drude model, dashed lines for the plasma model.   }
\end{figure}

In analogy with the decompostions of the forces in Eqs. (\ref{decomF}) and (\ref{decomF2}), we consider the following decompositions of the deviations from PFA:
\be
\beta_{n=0}=\frac{\tilde R}{a}\left(\frac{F_{n=0}}{F_{{\rm PFA} | n=0}}-1 \right)\;,
\ee   
\be
\beta_{n>0}=\frac{\tilde R}{a}\left(\frac{F_{n>0}}{F_{\rm PFA|n>0}}-1 \right)\;,
\ee 
\be
{\tilde \beta}_{n=0}=\frac{\tilde R}{a}\left(\frac{F'_{n=0}}{F'_{{\rm PFA} | n=0}}-1 \right)\;.
\ee 
and
\be
{\tilde \beta}_{n>0}=\frac{\tilde R}{a}\left(\frac{F'_{n>0}}{F'_{\rm PFA|n>0}}-1 \right)\;.
\ee 
The PFA values of the forces and force gradients that appear in the above equations are obtained by   the obvious modifications of the standard formulas Eq. (\ref{PFA}) and (\ref{PFA2}). For example, $F_{\rm PFA|n=0}=2 \pi {\tilde R} {\cal F}^{(\rm pp)}_{n=0}(a)$, while   $F_{\rm PFA|n>0}=2 \pi {\tilde R} {\cal F}^{(\rm pp)}_{n>0}(a)$. The corresponding quantities for the force gradient are defined in the same way. By a little computation one can prove the following two equations, which relate the full deviations to their two components:
\be
\beta= w\, \beta_{n=0}+ (1-w)\, \beta_{n>0}\;,\label{betacom}
\ee
\be
{\tilde \beta}= {\tilde w}\, {\tilde \beta}_{n=0}+ (1-{\tilde w})\, {\tilde \beta}_{n>0}\;,\label{tildebetacom}
\ee  
where the coefficients  $w$ and $\tilde w$ respectively represent the fractional contributions of the classical PFA terms   to the full PFA force and force gradient: 
\be
w(a)=\frac{F_{\rm PFA}\vert_{n=0}}{F_{\rm PFA}}\;,
\ee
\be
{\tilde w}(a)=\frac{F'_{\rm PFA}\vert_{n=0}}{F'_{\rm PFA}}\;.
\ee
We note that the coefficients $w(a)$ and ${\tilde w}(a)$ depend on the separation, but they are both independent of the spheres radii. The weights $w$ and $\tilde w$ are displayed in Fig. \ref{wplot} for the Drude model (solid lines) and for the plasma model (dashed lines).  As expected, the plots show that the weight of the classical term is rather small for small separations, while it becomes larger and larger as the separation increases.  The plot in Fig. \ref{wplot} shows also that, compared to the Drude mode,  the classical contribution has a larger weight within the plasma model, because within  the latter model both TE and TM polarizations  contribute to the classical term, while in the Drude model only the TM modes contribute for zero frequency. Keeping in mind Eqs. (\ref{betacom}) and (\ref{tildebetacom}), it is clear that for typical experimental submicron separations, the influence of the classical term gets suppressed by its relatively small weight $w$, and therefore one can predict that the magnitude of the total deviation from PFA   is  determined to a large extent by the positive Matsubara modes.       

\begin{figure}
\includegraphics[width=.9\columnwidth]{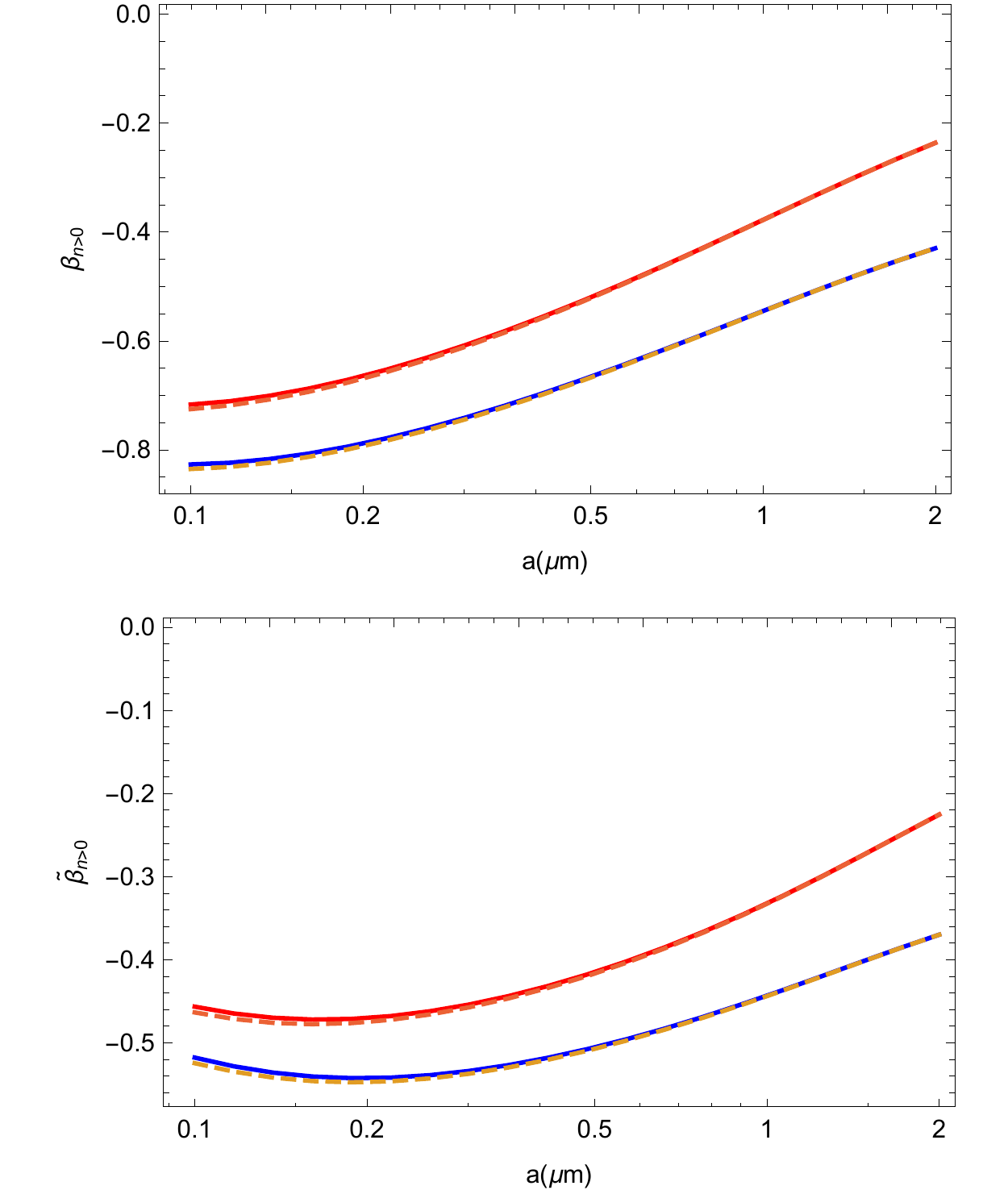}% Here is how to import EPS art
\caption{\label{betapos} Coefficients $\beta_{n>0}$ (upper panel) and ${\tilde \beta}_{n>0}$ (lower panel), computed using the DE. Solid and dashed lines correspond, respectively, to the Drude and plasma prescriptions. Red curves are for the sphere-plate geometry, blue curves for two spheres of equal radii.}
\end{figure}

We examine separately the contributions of the positive and classical  modes to $\beta$ and $\tilde \beta$. 
The values of the deviations $\beta_{n>0}$ and ${\tilde \beta}_{n>0}$ for the positive modes need not be computed, as they follow directly from  Eqs. (\ref{DEformula1}) and (\ref{DEformula2}):
\be
\beta_{n>0}= -\left({\theta}(a)+u \,{ \kappa}(a) \right) \;\label{betaDE}
\ee   
and
\be
{\tilde \beta}_{n>0}= -\left({\tilde \theta}(a)+u \,{\tilde \kappa}(a) \right) \;.\label{tildebetaDE}
\ee 
These equations show that within the DE the coefficients $\beta_{n>0}$ and $\tilde \beta_{n>0}$ depend on the separation and on the ratio among the radii, via the parameter $u$, but they are both independent of the effective radius $\tilde R$. We remark that within the DE there is no difference between  grounded and isolated conductors.  The values of the  coefficients $\beta_{n>0}$ and ${\tilde \beta}_{n>0}$, within the Drude and the plasma prescriptions, can be computed using the values of $(\theta,\kappa)$ and  $({\tilde \theta},{\tilde \kappa})$ listed in Tables I-IV.  Since the   latter are practically independent of the used prescription,  the values of the coefficients  $\beta_{n>0}$ and ${\tilde \beta}_{n>0}$ for the two prescriptions are practically coinciding. This expectation  is confirmed by Fig. \ref{betapos}, which shows plots of $\beta_{n>0}$ (upper panel) and ${\tilde \beta}_{n>0}$ (lower panel), computed using    the Drude prescription (solid lines) and the  plasma prescription (dashed lines). The red curves in  Fig. \ref{betapos} are for the sphere-plate geometry, while the blue curves are for two spheres of equal radii. According to Eqs. (\ref{betaDE}) and (\ref{tildebetaDE}) the curves corresponding to all other values of $R_1/R_2$ lie in the strip bounded by the sphere-plate and the $R_1=R_2$ curves of Fig. \ref{betapos}.  

\begin{figure}
\includegraphics[width=.9\columnwidth]{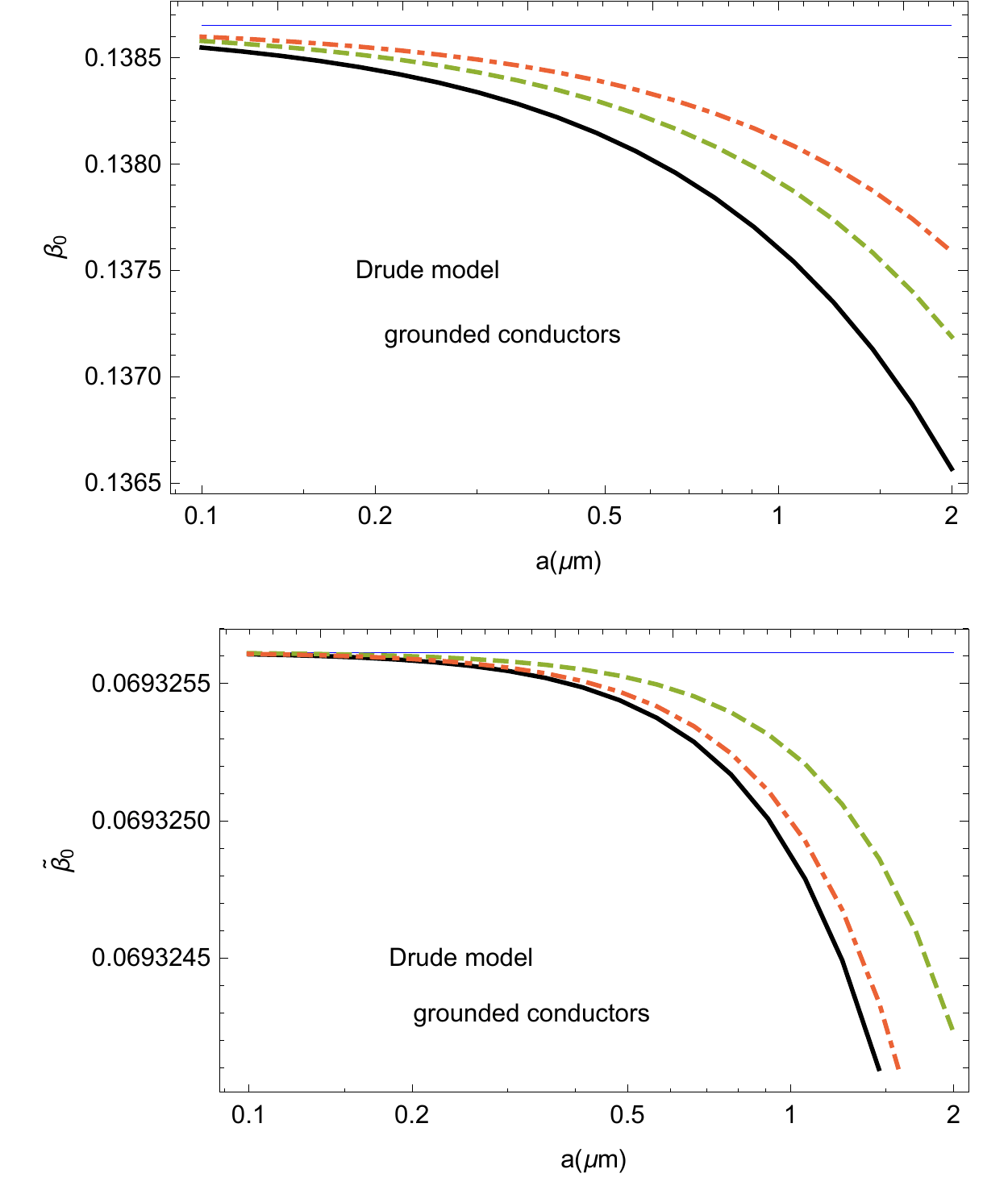}% Here is how to import EPS art
\caption{\label{beta0Dir} Coefficients $\beta_{n=0}$ (upper panel) and ${\tilde \beta}_{n=0}$ (lower panel) for the Drude model, with grounded conductors. Solid lines are for two identical spheres with radius $R=40\;\mu$m, dashed lines are for two spheres of radii $R_1=40\;\mu$m and  $R_2=80\;\mu$m, and the dot-dashed lines  are for two spheres of radii $R_1=40\;\mu$m and  $R_2=200\;\mu$m. The thin horizontal blue lines represents the prediction of the DE. }
\end{figure}

\begin{figure}
\includegraphics[width=.9\columnwidth]{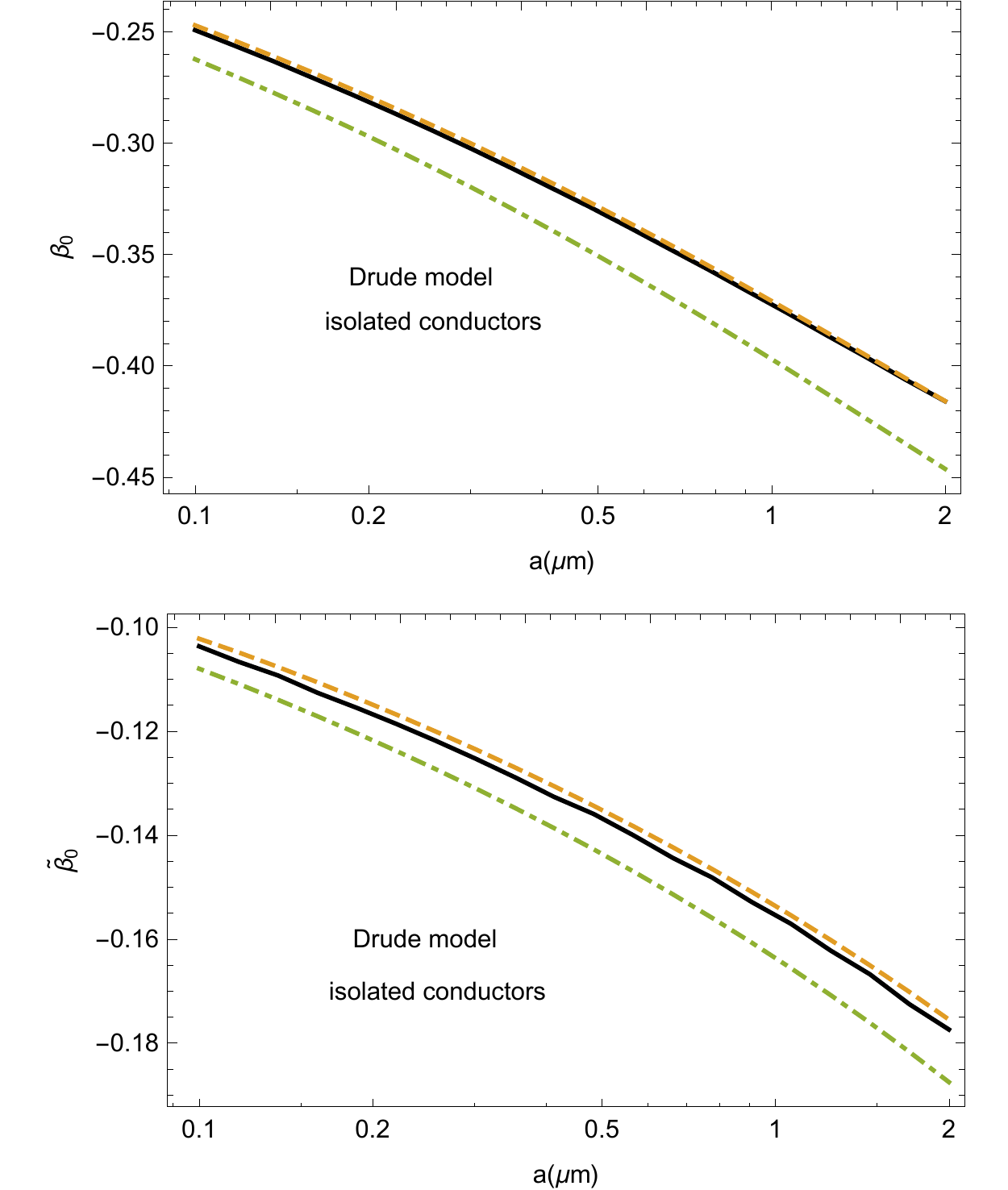}% Here is how to import EPS art
\caption{\label{beta0Drude} Coefficients $\beta_{n=0}$ (upper panel) and ${\tilde \beta}_{n=0}$ (lower panel) for the Drude model, with  isolated conductors. Solid lines are for two identical spheres with radius $R=40\;\mu$m, dashed lines are for two spheres of radii $R_1=40\;\mu$m and  $R_2=80\;\mu$m, and the dot-dashed lines  are for two spheres of radii $R_1=40\;\mu$m and  $R_2=200\;\mu$m.   }
\end{figure}

We pass now to the deviations from PFA for the classical force and force gradient. Plots of the coefficients $\beta_{n=0}$  and   ${\tilde \beta}_{n=0}$  for the Drude model are displayed in  Fig. \ref{beta0Dir} for  grounded conductors, and  in Fig. \ref{beta0Drude}  for isolated  conductors. Since the Drude bc for zero-frequency does not involve any length parameter besides the radii of the spheres and the separation, the coefficients $\beta_{n=0}$ and ${\tilde \beta}_{n=0}$  are indeed functions of the two dimensionless variables $a/{\tilde R}$ and $u$.  Therefore, it would be  natural to display both $\beta_{n=0}$ and ${\tilde \beta}_{n=0}$ as  functions of the dimensionless distance $a/{\tilde R}$. However, this would make a comparison with the plots of  $\beta_{n>0}$ and $\tilde \beta_{n>0}$  less straightforward, and for this reason we found preferable to display  $\beta_{n=0}$ and ${\tilde \beta}_{n=0}$ in Figs. \ref{beta0Dir} and \ref{beta0Drude}  as  functions of the dimensionful separation $a$.  
Using  Eq. (\ref{DEnzero}),  we see that the DE predicts for the Drude model  constant values for the deviations from PFA, i.e.  $\beta_{n=0}|_{\rm DE}=1/(6 \zeta(3))$ and  ${\tilde \beta}_{n=0}|_{\rm DE}=1/(12 \zeta(3))$,  independently of whether the conductors are grounded or not.   From  Fig. \ref{beta0Dir}, we see that  for grounded conductors ${\beta}_{n=0}$ and  ${\tilde \beta}_{n=0}$ are actually extremely close to the prediction of the DE (displayed by the thin horizontal blue lines in Fig. \ref{beta0Dir}). On the contrary, for isolated conductors, the numerical values of   $\beta_{n=0}$ and ${\tilde \beta}_{n=0}$ are  different from those predicted by the DE,  even in the sign. This by no means signifies that the DE fails in the latter case. The observed disagreement with the DE likely has the same explanation as in the sphere-plate case \cite{bimonteex1}, where it was found that for realistic separations the leading deviation from PFA  predicted by the DE  is actually dominated by large sub-leading logarithmic corrections.
\begin{figure}
\includegraphics[width=.9\columnwidth]{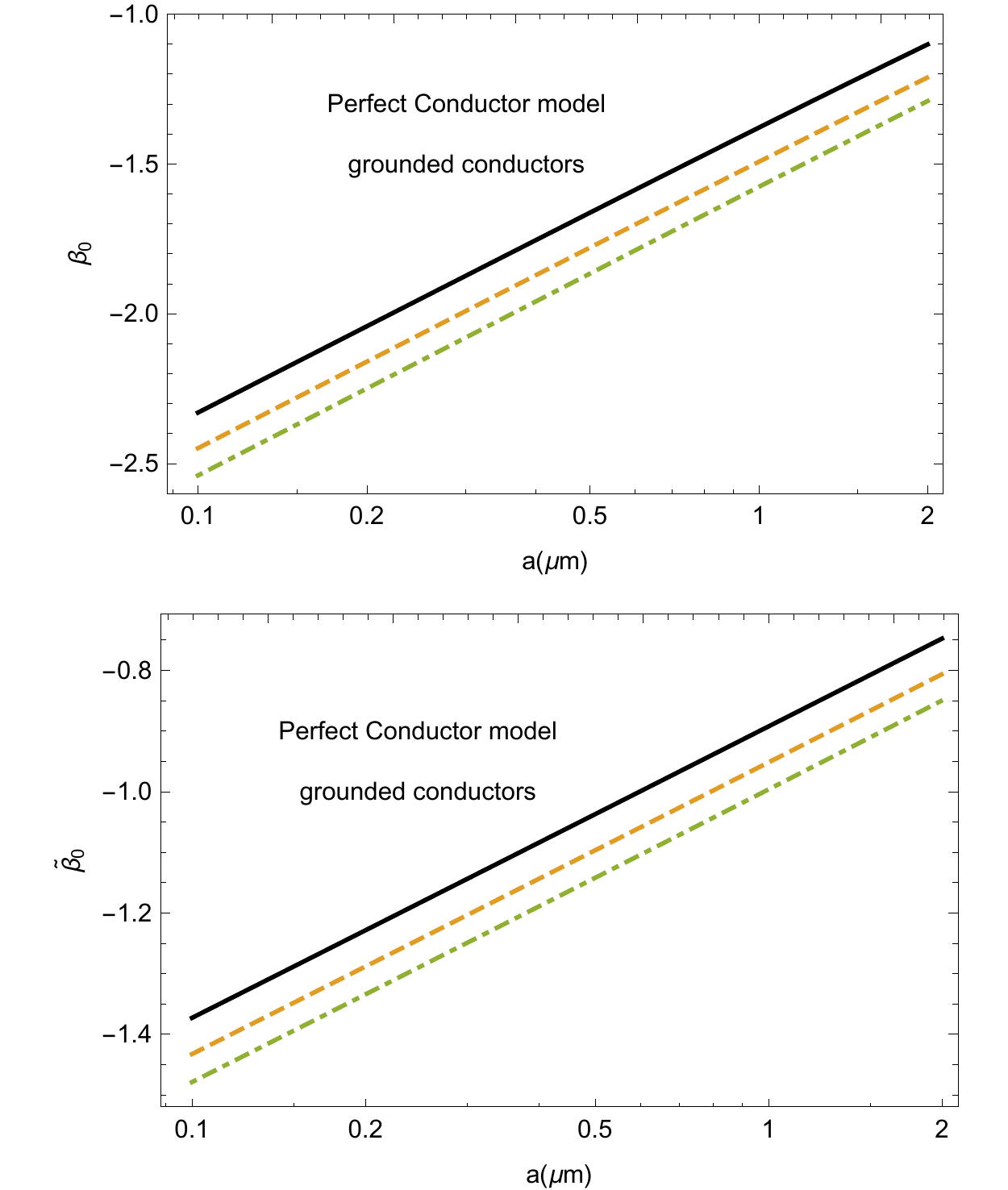}% Here is how to import EPS art
\caption{\label{beta0PCDir} Coefficients $\beta_{n=0}$ (upper panel) and ${\tilde \beta}_{n=0}$ (lower panel) for  perfect grounded  conductors. Solid lines are for two identical spheres with radius $R=40\;\mu$m, dashed lines are for two spheres of radii $R_1=40\;\mu$m and  $R_2=80\;\mu$m, and the dot-dashed lines  are for two spheres of radii $R_1=40\;\mu$m and  $R_2=200\;\mu$m.   }
\end{figure} 

\begin{figure}
\includegraphics[width=.9\columnwidth]{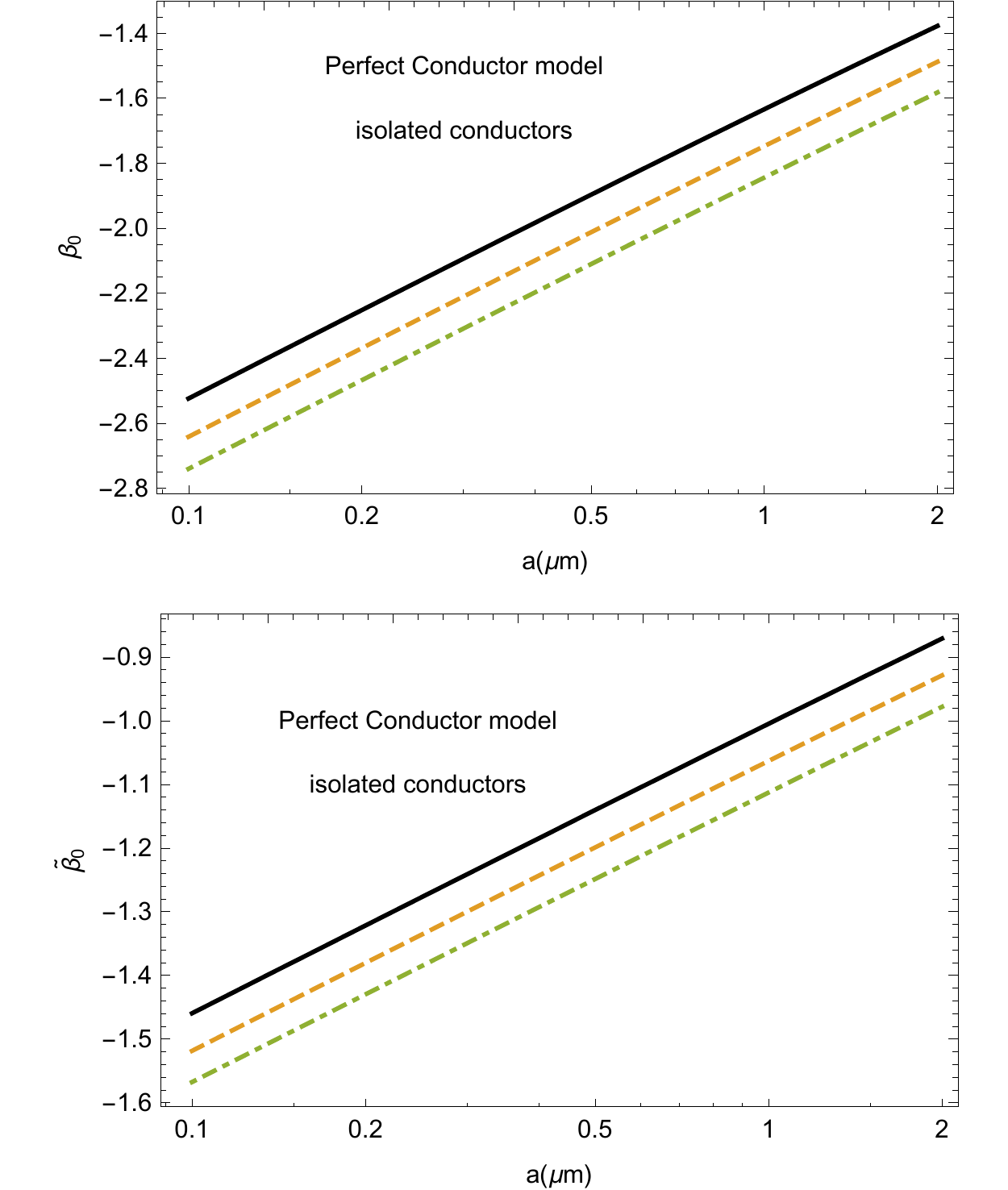}% Here is how to import EPS art
\caption{\label{beta0PCDr} Coefficients $\beta_{n=0}$ (upper panel) and ${\tilde \beta}_{n=0}$ (lower panel) for a perfect isolated conductor. Solid lines are for two identical spheres with radius $R=40\;\mu$m, dashed lines are for two spheres of radii $R_1=40\;\mu$m and  $R_2=80\;\mu$m, and the dot-dashed lines  are for two spheres of radii $R_1=40\;\mu$m and  $R_2=200\;\mu$m.   }
\end{figure}

In Figs. \ref{beta0PCDir} and \ref{beta0PCDr}  we plot $\beta_{n=0}$ (upper panel) and ${\tilde \beta}_{n=0}$ (lower panel) for the PC model, with grounded and isolated conductors respectively.  As we expalined in the previous section, the PC model coincides with the limit of the plasma model, for infinite plasma frequency. The striking feature displayed by  Figs.  \ref{beta0PCDir} and \ref{beta0PCDr}  is that the deviations from PFA for a PC have a much larger magnitude, compared to the Drude model. The cause of this "anomaly" is the zero-frequency TE mode, which for a PC is described by a  N scalar field, as we have seen in the previous section. 

Finally, we consider the plasma prescription. Within the latter model, the coefficients ${\beta}_{n=0}$ and  ${\tilde \beta}_{n=0}$ depend parametrically on the plasma frequency $\omega_p$ of the conductors, or better on the plasma length $\lambda_p=c/\omega_p$ ($\lambda_p=22$ nm for gold).     It turns out that already for separations $a$ larger than a few times $\lambda_p$,  ${\beta}_{n=0}$ and ${\tilde  \beta}_{n=0}$ are close to the PC limit. This can be seen from Table \ref{tab.5}, which compares the values of $\beta_{n=0}$ for two equal isolated spheres of radius $R$ = 10 $\mu$m, computed using  the plasma model with those obtained by  the PC model. The difference between the two models reaches a maximum of 8 percent, for the smallest separation $a=100$ nm, and decreases quickly as the separation increases. This suggests that one could well use $\beta_{n=0}^{(\rm PC)}$ in the place of  $\beta_{n=0}^{(\rm pl)}$ in Eq. (\ref{betacom}), to compute $\beta$ for the plasma model. We remind that  $\beta_{n=0}^{(\rm PC)}$ can be computed quickly and to high precision using bispherical coordinates. The  substitution  of   $\beta_{n=0}^{(\rm pl)}$ by  $\beta_{n=0}^{(\rm PC)}$ indeed engenders  small errors on the values of $\beta$, as  shown  by the last two rows of Table \ref{tab.5}, where we compare  the exact plasma values of $\beta^{(\rm pl)}$ with the approximate values 
$\beta^{(\rm pl)}_{\rm app}$ that are obtained in this way. The maximum error  of 1.1 percent occurs for the smallest considered separation of 100 nm.  Please note however that a 1.1 \% error on $\beta$ for $a=100$ nm, signifies an error ${\tilde R}/a=50$ times smaller on the Casimir force, i.e. an error of $2 \times 10^{-4}$ on the force, which is far beyond the present accuracy of Casimir experiments. The error on $\beta$   decreases to 0.5 \% already for 200 nm, and  becomes totally negligible for separations larger than 300 nm.  These considerations justify using the perfect conductor model to estimate $\beta_{n=0}$ within the plasma prescription.
\begin{table}[h]
\begin{tabular}{ccccccc} \hline 
$a (\mu m)$\;\; &0.1&\;\;0.2\;\;&0.3\;\;&\;0.4 \;\;& \;\;\;0.9\;\;& \;\;2\;\; \\ \hline \hline
${\beta}_{n=0}^{(\rm pl)}$\;\;&-1.82&\;\; -1.66\;\; &-1.54\;\;&-1.44 &\;\; -1.16\;\; &\;\; -0.89    \\  \hline
${\beta}_{n=0}^{(\rm PC)}$\;\;&-1.98&\;\; -1.72\;\; &-1.57\;\;& -1.46 &\;\; -1.17\;\; &\;\; -0.89    \\ \hline
${\beta}^{(\rm pl)}$\;\;&-0.896&\;\; -0.889\;\; &-0.869\;\;& -0.852 &\;\; -0.817\;\; &\;\; -0.796    \\  \hline
${\beta}_{\rm app}^{(\rm pl)}$\;\;&-0.906&\;\; -0.894\;\; &-0.873\;\;& -0.855 &\;\; -0.819\;\; &\;\; -0.796    \\ \hline 
\end{tabular}
\caption{Values of the coefficients ${\beta}_{n=0}$ for two equal isolated spheres of radius $R$ = 10 $\mu$m,  computed using the  plasma model ($\omega_{\rm p}=9$ eV/$\hbar$) or the perfect conductor model. The fourth line of the Table displays the values of $\beta^{(\rm pl)}$, while the fifth line displays the  approximate values of $\beta^{(\rm pl)}$ that result from substituting $\beta_{n=0}^{(\rm pl)}$ by $\beta_{n=0}^{(\rm PC)}$ in Eq. (\ref{betacom}).}
\label{tab.5}
\end{table}

\begin{figure}
\includegraphics[width=.9\columnwidth]{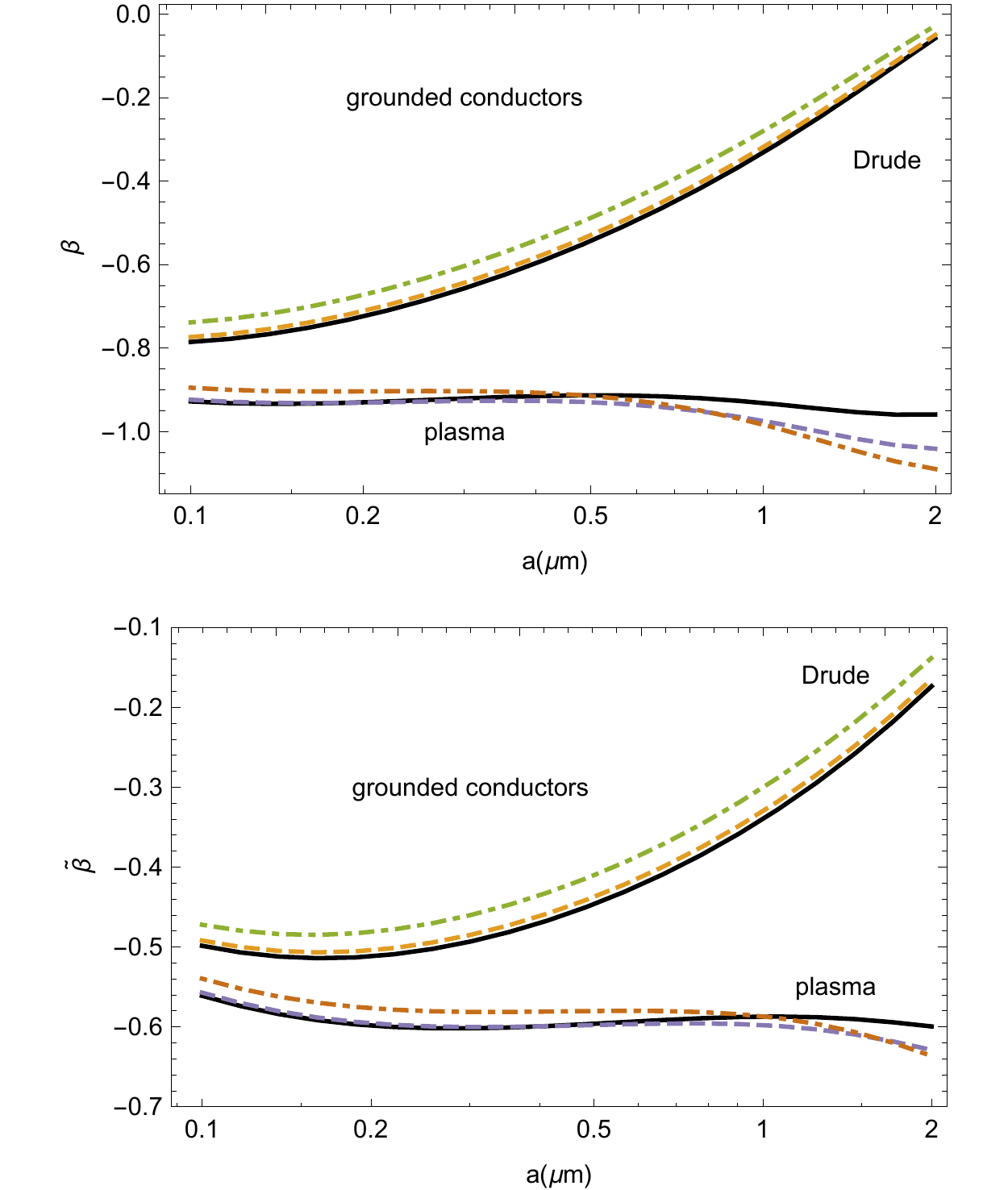}% Here is how to import EPS art
\caption{\label{betaDirpl}  Coefficients $\beta$ (upper panel) and ${\tilde \beta}$ (lower panel) for grounded  conductors.   Solid lines are for two identical spheres with radius $R=40\;\mu$m, dashed lines are for two spheres of radii $R_1=40\;\mu$m and  $R_2=80\;\mu$m, and the dot-dashed lines  are for two spheres of radii $R_1=40\;\mu$m and  $R_2=200\;\mu$m.   }
\end{figure}

\begin{figure}
\includegraphics[width=.9\columnwidth]{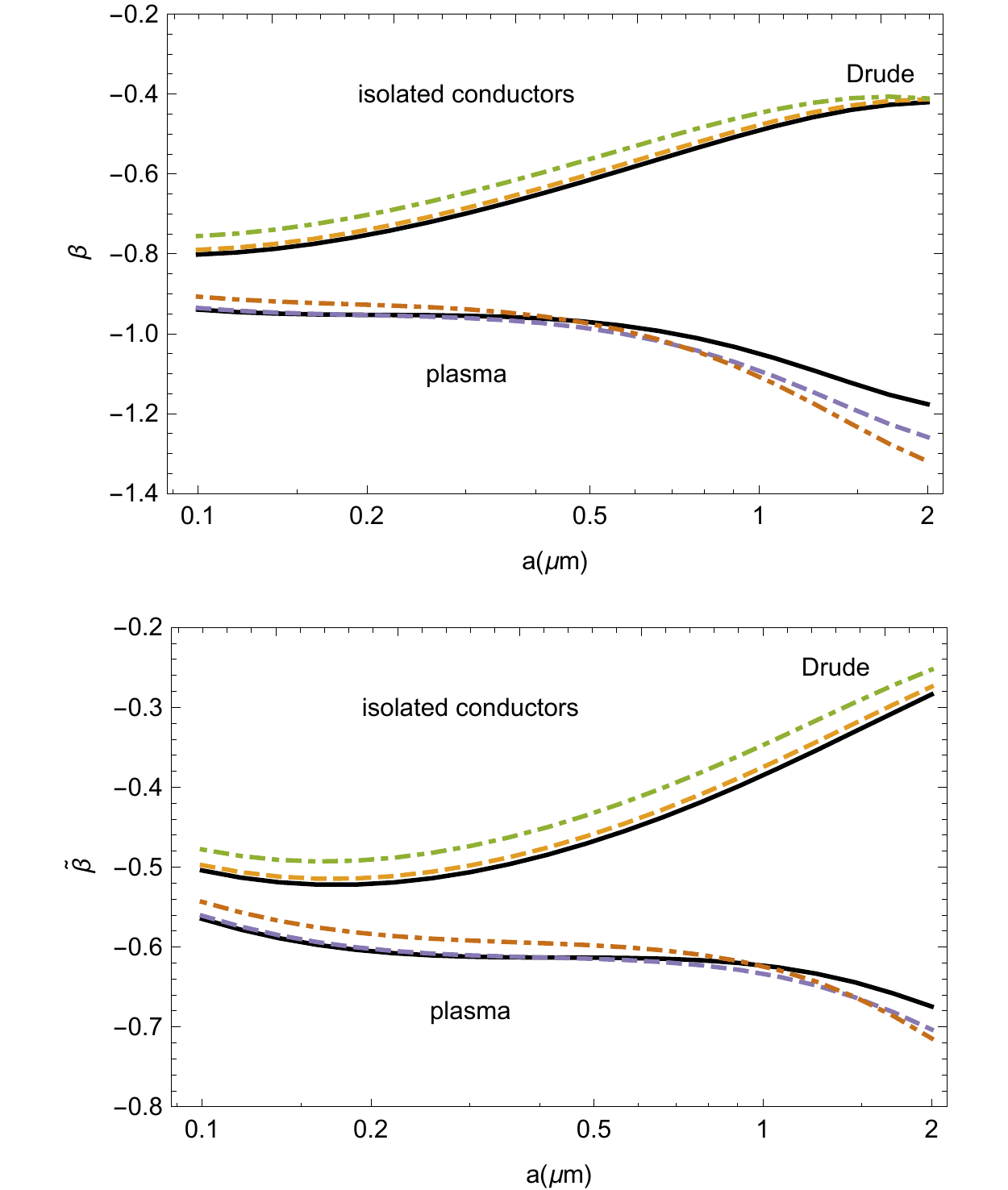}% Here is how to import EPS art
\caption{\label{betaDrpl}  Coefficients $\beta$ (upper panel) and ${\tilde \beta}$ (lower panel), for isolated conductors.  Solid lines are for two identical spheres with radius $R=40\;\mu$m, dashed lines are for two spheres of radii $R_1=40\;\mu$m and  $R_2=80\;\mu$m, and the dot-dashed lines  are for two spheres of radii $R_1=40\;\mu$m and  $R_2=200\;\mu$m.   }
\end{figure}

Now, we are ready to present the full coefficients $\beta$ and $\tilde \beta$ for the above models of the conductors. In Fig. \ref{betaDirpl} we display these coefficients for grounded conductors, while in Fig. \ref{betaDrpl} the same coefficients are displayed for isolated  conductors. Both figures show that for all considered separations and for all models the deviation from PFA is of order one. We also note that the magnitude of the deviations from PFA engendered by the plasma model are always larger than those of the Drude model, in qualitative agreement with the findings of \cite{gert}. In fact, the increased magnitude of the deviations from PFA for the  plasma model  is manifest  for  separations larger than one micron, and is particularly evident  for grounded conductors.  For $R_2 \gg R_1$, the deviations from PFA obtained by us for the two-sphere system, for both the Drude and the plasma models with isolated  conductors (the only case studied in \cite{gert}),   reproduce the corresponding  deviations for the sphere-plate geometry that were computed in \cite{gert}. 

\begin{figure}
\includegraphics[width=.9\columnwidth]{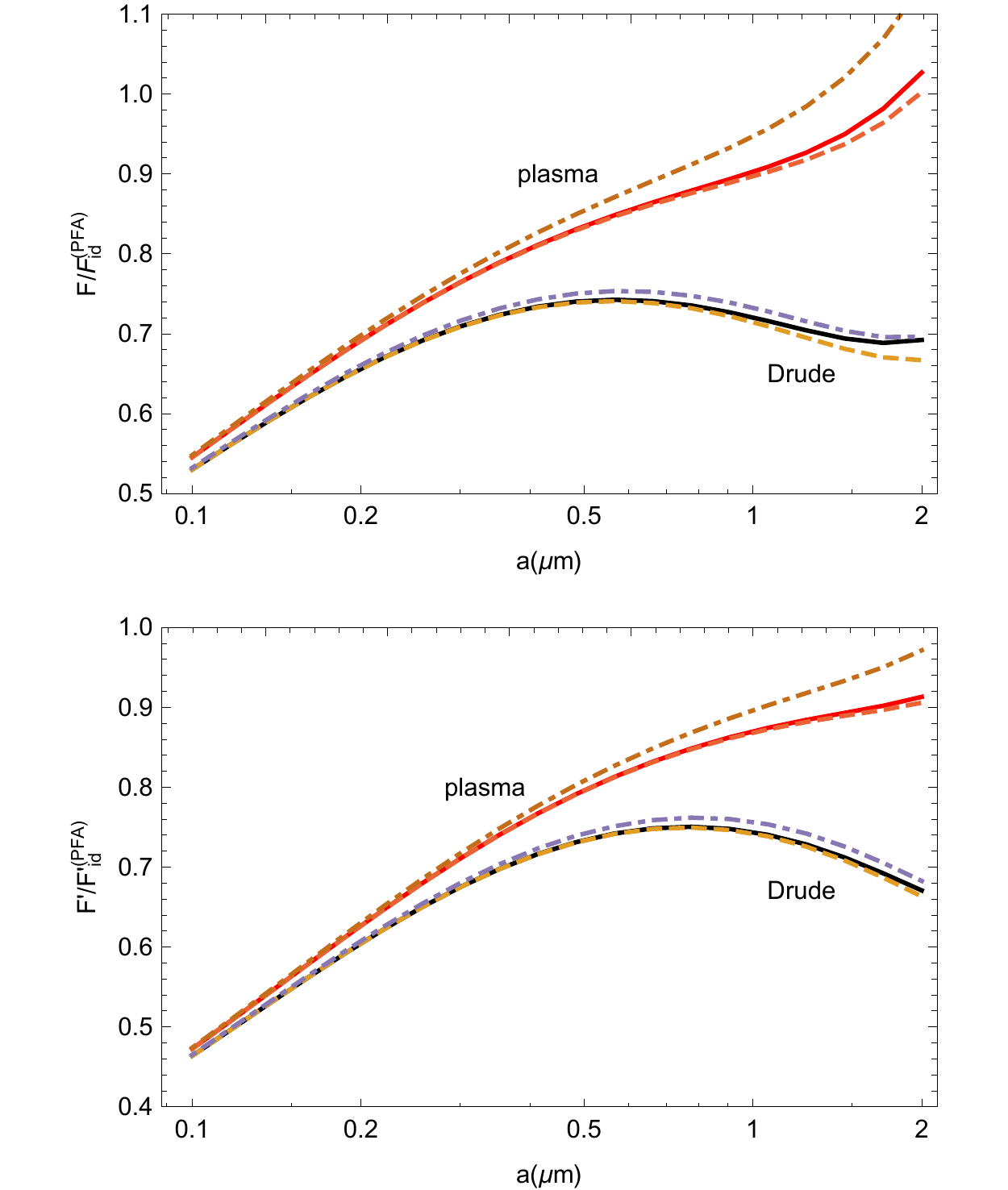}% Here is how to import EPS art
\caption{\label{normfor} Room temperature Casimir force between two identical gold sphere of radius $R=40$ $\mu$m,  normalized by the PFA force $F^{\rm (id)}_{\rm PFA}$ for ideal spheres at zero temperature  (upper panel), and  the corresponding force gradient  normalized by the force gradient  $F'^{\rm (id)}_{\rm PFA}$  for ideal spheres at zero temperature (lower panel).   Solid lines are for grounded  spheres,  dashed lines are for isolated spheres, while dot-dashed lines represent the standard PFA.}
\end{figure}

In Fig. \ref{normfor} we display the room temperature Casimir force (upper panel) and the force gradient (lower panel) for two identical gold spheres of radius $R=40$ $\mu$m. The Casimir  force is normalized by the PFA force $F^{(\rm id)}_{\rm PFA}=-\pi^3 \hbar c {\tilde R}/(360 a^3) $  for two ideal spheres at zero temperature, while the force gradient is normalized by the corresponding ideal force gradient  $F'^{(\rm id)}_{\rm PFA}=\pi^3 \hbar c {\tilde R}/(120 a^4) $. The solid lines in Fig. \ref{normfor} are for grounded spheres,  the dashed lines are for isolated spheres, while the dot-dashed lines represent the ordinary PFA forces. The figure shows clearly that for separations larger than 500 nm, the plasma forces deviate from the PFA much more than the corresponding forces for the Drude model. The figure shows also that  grounding the conductors leads to a very  small correction to the forces for submicron separations, but its effect becomes visible for large separations.  
 
The experiment \cite{garrett} measured the Casimir force gradient between two gold-coated spheres at room temperature  in vacuum.  A fit of the data taken with spheres of different radii in the separation interval from 40 to 300 nm, was used to set the bound ${\beta'}=-6 \pm 27$  on the deviations from PFA of the Casimir force gradient. The quantity $\beta'$ of \cite{garrett} is the same as the coefficient $\tilde \beta$ of the present work, and thus we see from Figs. \ref{betaDirpl} and \ref{betaDrpl} that the bound is consistent with the theoretical  predictions of all models considered in this paper. A discrimination between the deviations from PFA predicted by these models  requires a significant improvement, by one or two orders of magnitude, in the sensitivity of the experimental apparatus used in \cite{garrett}.

\section{Conclusions}

We have computed the deviations from PFA for the Casimir force and force-gradient between two sphere in vacuum, at room temperature. The computations have been carried out for four distinct models of the conductors, i.e. the Drude and plasma models, with grounded or isolated  conductors. We notice that, while the Drude and plasma models have been investigated in many studies before, but not in the sphere-sphere geometry considered in this work, only rarely the effect of grounding the  conductors has been discussed in the literature.  Despite the fact that all Casimir experiments utilize  conductors connected to charge reservoirs, the standard theoretical models used to interpret the experiments exclude from the start fluctuations of the total charges of the  conductors, and thus describe isolated  conductors. Probably, the influence of grounding the  conductors has escaped detection so far because experiments are carried out in the small separation regime, where the effects of grounding are negligible, since they become manifest only at the level of deviations from PFA.   
 
 For all models considered in this work, the magnitudes of the obtained theoretical  deviations from PFA  are of order one, and thus they are all  consistent with the loose experimental bound ${\tilde\beta}=6 \pm 27$ that was set by the recent experiment \cite{garrett}, which measured the force-gradient between two gold coated spheres in vacuum. Our results indicate that a significant improvement in the sensitivity of the apparatus, by one or two orders of magnitude, is necessary in order  to discriminate between the deviations from PFA predicted by the four theoretical models.

\acknowledgments

The author thanks T. Emig, N. Graham, M. Kr\"uger, R. L. Jaffe and M. Kardar for valuable discussions.  The author acknowledges an anonymous referee for suggesting that  a certain amount of computing time could be saved  by using zeta function/heat kernel   tecniques (along the lines of Ref. \cite{asorey})  to handle the sum over Matsubara frequencies.


\newpage

\begin{thebibliography}{99}



\bibitem{Casimir48} H.~B.~G. Casimir, Proc. K. Ned. Akad. Wet., {\bf 51},  793 (1948).

\bibitem{book1} K. A. Milton,  The Casimir Effect: Physical manifestations of Zero-Point Energy, World Scientific, Singapore (2001).



\bibitem{parse} V. A. Parsegian, Van der Waals Forces, Cambridge University Press (2005).

\bibitem{book2} M. Bordag, G. L. Klimchitskaya, U. Mohideen and V. M. Mostepanenko,  Advances in the Casimir Effect, Oxford University Press (2009).

\bibitem{buhmann} S. Y. Buhmann, {\it Dispesion forces} (Springer, Heidelberg, 2012).

\bibitem{woods} L. M. Woods, D.A.R.  Dalvit, A. Tkatchenko, P. Rodriguez-Lopez, A.W. Rodriguez, and R. Podgornik, Rev. Mod. Phys. {\bf 88}, 045003 (2016).

\bibitem{mehran}  G. Bimonte, T. Emig, M. Kardar, and M. Kr\"uger, Ann. Rev. Cond. Matt. Phys. {\bf 8}, 119 (2017).

\bibitem{decca6} R. S. Decca, D.  L\'{o}pez, E. Fischbach et al, Eur. Phys. J. C {\bf 51},  963 (2007).

\bibitem{decca3} Y.-J. Chen, W. K. Tham, D. E. Krause, D. Lopez, E. Fischbach, R. S. Decca, Phys. Rev. Lett. {\bf 116}, 221102 (2016).

\bibitem{Derjaguin} B. Derjaguin, Kolloid Z. {\bf 69}, 155 (1934).

\bibitem{lifs} E. M. Lifshitz, Zh. Eksp. Teor. Fiz. {\bf 29}, 94 (1955) [Sov. Phys. JETP {\bf 2}, 73 (1956)].

\bibitem{deccamag} G. Bimonte, D. L\'opez, and R. S. Decca,  Phys. Rev. B 93, 184434 (2016).

\bibitem{Balian} R. Balian, B. Duplantier, Ann. Phys. {\bf 104}, 300 (1977); {\it ibid.} {\bf 112}, 165 (1978).

\bibitem{langbein} D. Langbein,Theory of van der Waals attraction, Springer (1974).

\bibitem{sca1} A. Lambrecht, P. A. Maia Neto, and S. Reynaud,  New J. Phys.
{\bf 8}, 243 (2006).

\bibitem{sca2} T. Emig, N. Graham, R. L. Jaffe, and M. Kardar, Phys. Rev. Lett. {\bf 99}, 170403 (2007).

\bibitem{kenneth} O. Kenneth and I. Klich, Phys. Rev. Lett. {\bf 97}, 160401 (2006); Phys. Rev. B {\bf 78}, 014103 (2008).

\bibitem{bordag} M. Bordag, Phys. Rev. D {\bf 73}, 125018 (2006).

\bibitem{paulo} P.A. Maia Neto, A. Lambrecht, and S. Reynaud, Phys. Rev. A {\bf 72}, 012115 (2005).

\bibitem{fosco1} C. D. Fosco, F. C. Lombardo, and F. D. Mazzitelli,  Phys. Rev. D {\bf 84}, 105031 (2011).

\bibitem{bordagteo} L. P. Teo, M. Bordag, and V. Nikolaev, Phys. Rev. D {\bf 84}, 125037 (2011).

\bibitem{bimonte1} G. Bimonte, T. Emig, R.L. Jaffe, and M. Kardar,  EPL {\bf 97}, 50001 (2012).

\bibitem{fosco2} C. D. Fosco, F. C. Lombardo, and F. D. Mazzitelli,   Phys. Rev. D {\bf 86}, 045021 (2012).

\bibitem{fosco3}  C. D. Fosco, F. C. Lombardo, and F. D. Mazzitelli,   Phys. Rev. A {\bf 89}, 062120 (2014).

\bibitem{bimonte2} G. Bimonte, T. Emig,   and M. Kardar,  Appl. Phys. Lett. {\bf 100} 074110 (2012).

\bibitem{antoine1} A. Canaguier-Durand, P. A. Maia Neto, I. Cavero-Pelaez, A. Lambrecht, and S. Reynaud,
Phys. Rev. Lett. {\bf 102}, 230404 (2009).

\bibitem{antoine2} A. Canaguier-Durand, P. A. Maia Neto,  A. Lambrecht, and S. Reynaud,
Phys. Rev. A {\bf 82}, 012511 (2010).


\bibitem{gert} M. Hartmann, G.-L. Ingold, and P. A. Maia Neto
Phys. Rev. Lett. {\bf 119}, 043901 (2017). 

\bibitem{ricardo} D. E. Krause, R. S. Decca, D. L\'opez, and E. Fischbach,  Phys. Rev. Lett. 98, 050403 (2007).

\bibitem{bimonteprecise} G. Bimonte, EPL {\bf 118}, 20002 (2017).

\bibitem{bimonteex1} G. Bimonte, T. Emig,   Phys. Rev. Lett. {\bf 109}, 160403 (2012).

\bibitem{garrett} J. L. Garrett, D. A. T. Somers, and J. N. Munday, Phys. Rev. Lett. {\bf 120}, 040401 (2018).

\bibitem{pablo} P. Rodriguez-Lopez, Phys. Rev. B {\bf 84}, 075431 (2011).

\bibitem{teo2} L. P. Teo Phys. Rev. D {\bf 85}, 045027 (2012).

\bibitem{bimonte2sphere} G. Bimonte,  Phys. Rev. D {\bf 97}, 085011 (2018).

\bibitem{teo2s} L. P. Teo, Phys. Rev. D {\bf 90}, 045012 (2014).

\bibitem{deccapat} R. O. Behunin, D. A. R. Dalvit, R. S. Decca, C. Genet, I. W. Jung, A. Lambrecht, A. Liscio, D.  L\'opez, S. Reynaud, G. Schnoering, G. Voisin, and Y. Zeng, Phys. Rev. A {\bf 90},  062115 (2014).

\bibitem{foscogr}  C. D. Fosco, F. C. Lombardo, and F. D. Mazzitelli, Phys. Rev. D {\bf 94}, 085024 (2016).


\bibitem{fosco4} C. D. Fosco, F. C. Lombardo, and F. D. Mazzitelli,   Phys. Rev. D {\bf 92}, 125007 (2015).

\bibitem{bimonteHT} G. Bimonte, Phys. Rev. D {\bf 95}, 065004 (2017).



\bibitem{rahi} S. J. Rahi, T. Emig, N. Graham, R. L. Jaffe, and M. Kardar,
Phys. Rev. D {\bf 80}, 085021 (2009).



\bibitem{palik}  {\it Handbook of Optical Constants of Solids},
edited by E. D. Palik (Academic, New York, 1995).

\bibitem{bimonteKK} G. Bimonte, {\it Phys. Rev. A}, {\bf 83} (2011) 042109.

\bibitem{morse} P.M. Morse and H. Feshbach, {\it Methods of Theoretical Physics} (McGraw-Hill, New York, 1953), Part II, p. 1298.



\bibitem{antoine} A. Canaguier-Durand, G.-L. Ingold, M.-T. Jaekel, A. Lambrecht, P. A. Maia Neto, and S. Reynaud, Phys. Rev. A {\bf 85}, 052501 (2012).

\bibitem{asorey} M. Asorey and J. M. Mu$\tilde {\rm n}$oz-Casta$\tilde {\rm n}$eda, Nucl. Phys. B {\bf 874}, 852 (2013).

 














\end{thebibliography}
\end{document}